# A Comprehensive Review of U-Net and Its Variants: Advances and Applications in Medical Image Segmentation

Wang Jiangtao[1,2], Nur Intan Raihana Ruhaiyem[1], Fu Panpan[1]

1School of Computer Sciences, Universiti Sains Malaysia, Penang,11800, Malaysia
2 School of Network Communication, Zhejiang Yuexiu University, Shaoxing,312000, China

**Abstract:** Medical images often exhibit low and blurred contrast between lesions and surrounding tissues, with considerable variation in lesion edges and shapes even within the same disease, leading to significant challenges in segmentation. Therefore, precise segmentation of lesions has become an essential prerequisite for patient condition assessment and formulation of treatment plans. Significant achievements have been made in research related to the U-Net model in recent years. It improves segmentation performance and is extensively applied in the semantic segmentation of medical images to offer technical support for consistent quantitative lesion analysis methods. First, this paper classifies medical image datasets on the basis of their imaging modalities and then examines U-Net and its various improvement models from the perspective of structural modifications. The research objectives, innovative designs, and limitations of each approach are discussed in detail. Second, we summarize the four central improvement mechanisms of the U-Net and U-Net variant algorithms: the jump-connection mechanism, residual-connection mechanism, 3D-UNet, and transformer mechanism. Finally, we examine the relationships among the four core enhancement mechanisms and commonly utilized medical datasets and propose potential avenues and strategies for future advancements. This paper provides a systematic summary and reference for researchers in related fields, and we look forward to designing more efficient and stable medical image segmentation network models based on the U-Net network.

## 1. Introduction

Computer vision has greatly increased the intelligence level of medical image processing, reduced the intelligence level of medical staff, and improved the quality and efficiency of medical services. Simultaneously, it establishes a robust technical foundation for advancing intelligent healthcare and telemedicine to higher levels in the future. In terms of image analysis and recognition, whether traditional X-rays, CT scans, MRI, or more advanced PET and optical imaging, computer vision technology can automatically recognize and analyze the structure and features in medical images, such as identifying key information such as lesions, organ contours, and vascular distributions, which can help doctors quickly locate and judge possible lesions.

Convolutional neural networks (CNNs) have the potential to support clinicians in the early screening and diagnosis of diseases, thereby enhancing both detection rates and diagnostic accuracy for conditions such as cancer, cardiovascular diseases, and brain lesions. J et al. designed a fully convolutional neural network (FCN) [1]. Compared with CNNs, the most prominent feature is the use of convolutional layers instead of fully connected layers. The FCN represents a pioneering model for achieving pixel-level semantic segmentation of images. It has become an important framework for constructing semantic segmentation models since its proposal.

Table 1 Common medical image datasets and applicable network models

| Image Type | Dataset | Dataset Introduction | Suitable Model |
|---|---|---|---|
| Computed Tomography (CT) | LIDC_IDRI<br>LiTS<br>KiTS | lung nodule segmentation and detection<br>Liver and liver tumor segmentation<br>Targets kidney and kidney tumor segmentation | 3D U-Net[2]<br>V-Net[3]<br>HD$^2$A-Net[4]<br>DenseNet+U-Net[5] |
| Magnetic Resonance Imaging (MRI) | BraTS<br>ACDC<br>PROMISE12 | Specializes in brain tumor segmentation with multi-modal MRI data<br>Used for segmenting cardiac structures in MRI<br>Prostate segmentation in MRI images | 3D U-Net [2]<br>Attention U-Net [6]<br>nnU-Net [7]<br>DLF[8]<br>3DUV-NetR+[9] |
| Ultrasound | CAMUS<br>BUSI | Cardiac Acquisitions for Multi-structure Ultrasound Segmentation<br>Breast Ultrasound Images | U-Net[10]<br>H-TUNet[11]<br>DeepLabV3+[12] |
| X-Ray | Chest X-ray<br>Montgomery and Shenzhen []Chest X-ray Sets | Detecting and segmenting lung conditions<br>Primarily used for tuberculosis screening | U-Net [10]<br>FCN [1]<br>Attention U-Net [6] |
| Positron Emission Tomography (PET) | MSD<br>PET-CT | Organ segmentation from PET scans<br>Combines PET/CT fusion images, particularly in oncology | 3D U-Net[2]<br>DenseNet[13]<br>nnU-Net[7] |
| Optical Coherence Tomography (OCT) | Retina OCT<br>DUKE OCT<br>DRIVE | Retinal layer segmentation and disease detection<br>Retinal segmentation<br>Digital Retinal Images for Vessel Extraction | U-Net [10]<br>3D U-Net [2]<br>ResMUNet[14]<br>Attention U-Net[6] |
| Histopathology | CAMELYON16<br>PANDA | Detecting metastases in lymph node sections<br>Prostate cancer grading based on histopathology images. | U-Net [10]<br>Attention U-Net [6]<br>FCN[1]<br>DeepLabV3+[12] |

In the field of medical imaging, the diversity of imaging modalities has resulted in a wide range of datasets, each exhibiting distinct characteristics. This diversity leads to variations in the performance of different deep network models across these datasets. This study systematically reviews seven types of medical imaging datasets, each corresponding to a different imaging modality, and analyzes the deep learning models that are most suitable for each dataset. Detailed information is provided in Table 1. The provided data clearly indicate that U-Net and its variants have been extensively applied in medical image segmentation, establishing themselves as the predominant methodologies in this domain. Ronneberger et al. improved the FCN network and proposed a U-shaped network architecture, later known as U-Net[10]. The specific information is shown in Figure 1. The encoder of U-Net extracts image features through multiple convolution and pooling operations, whereas the decoder gradually restores feature images through upsampling. Finally, image segmentation is achieved through 1x1 convolutional mapping. The U-Net model is capable of being trained on small datasets while achieving outstanding performance, a characteristic that is particularly critical in medical image segmentation tasks. Medical datasets are typically limited in size, and the requirements for annotation are relatively high, which makes the generation of large-scale labeled datasets challenging. Consequently, various U-Net-based modified network models have received widespread attention in medical image segmentation research.

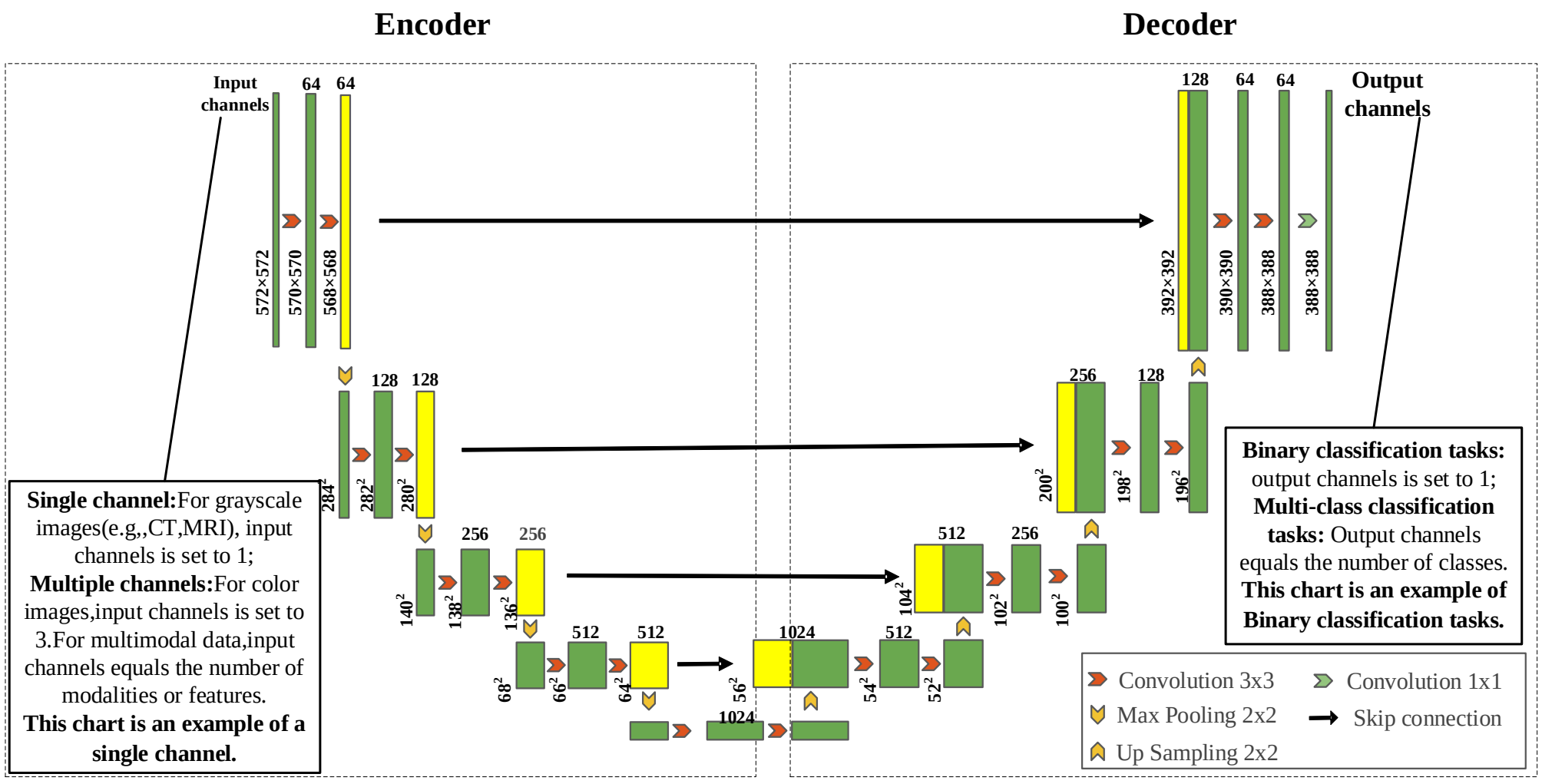

Figure 1. The Network Structure of U-Net

## 2. Diversified forms of U-Net and its variants

The U-Net model has demonstrated excellent performance in medical image segmentation, leading to the continuous development and application of various U-Net variants. There are three types of methods: the first approach involves optimizing the number of encoders; the second focuses on constructing multiple U-Net networks; and the third aims to develop 3D U-Net models for segmenting images with more complex spatial features. Some representative U-Net variants are depicted in Figure 2.

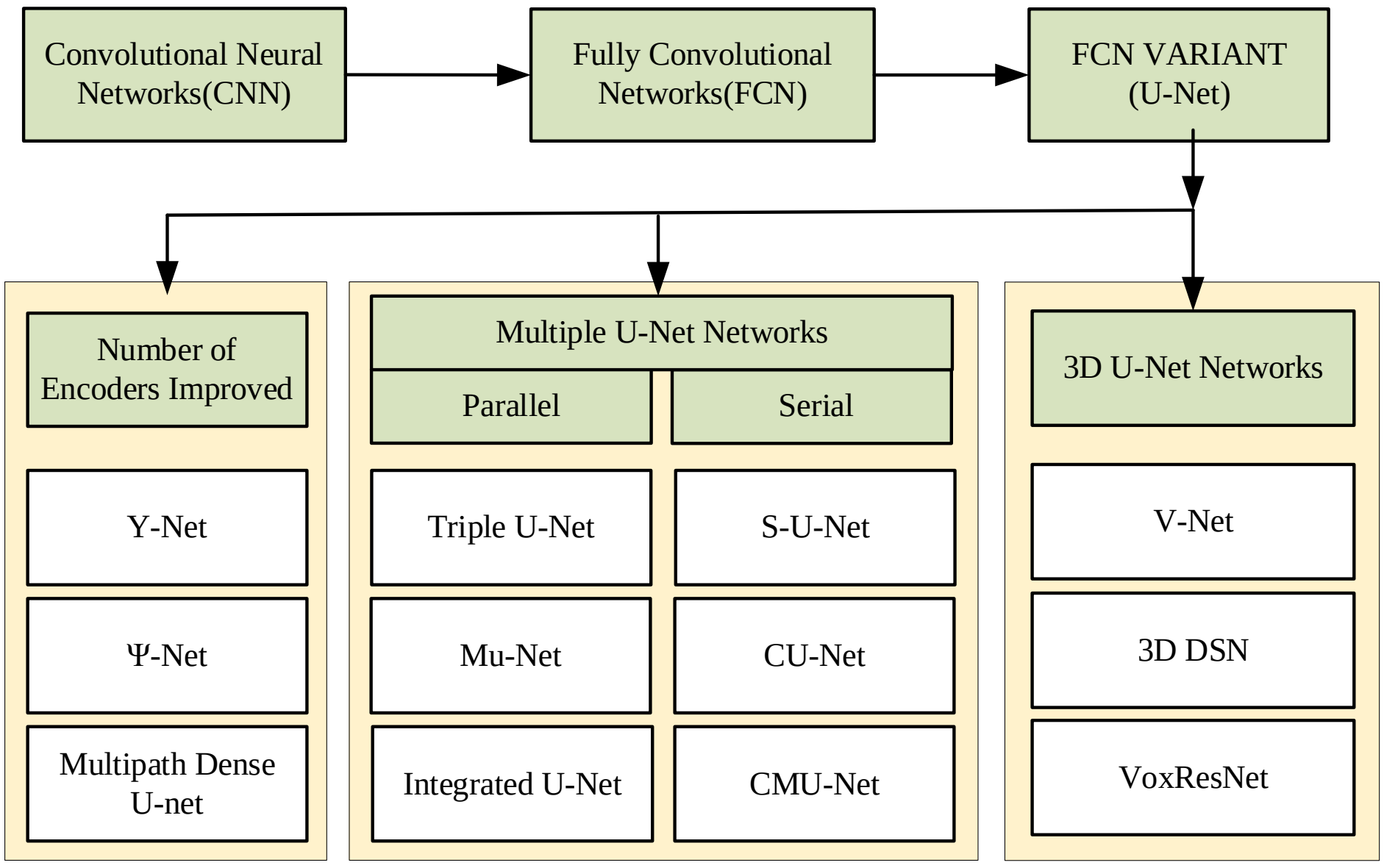

Figure 2. Diversified forms of U-Net and its variants

### 2.1 Improvement in the number of encoders

The image features are extracted by the encoder, while the decoder reconstructs the extracted feature map back to its original size. Significant improvements in the encoder architecture are found in Y-Net, Ψ-Net, and multipath dense U-Net. Lan et al. proposed the Y-Net [15] network, which solves the problem of mapping inverse models for complex targets in some special image processing tasks, such as reconstructing the initial optical pressure distribution for the original PA (photoacoustic) signal or beamformed image, as shown in Figure 3.

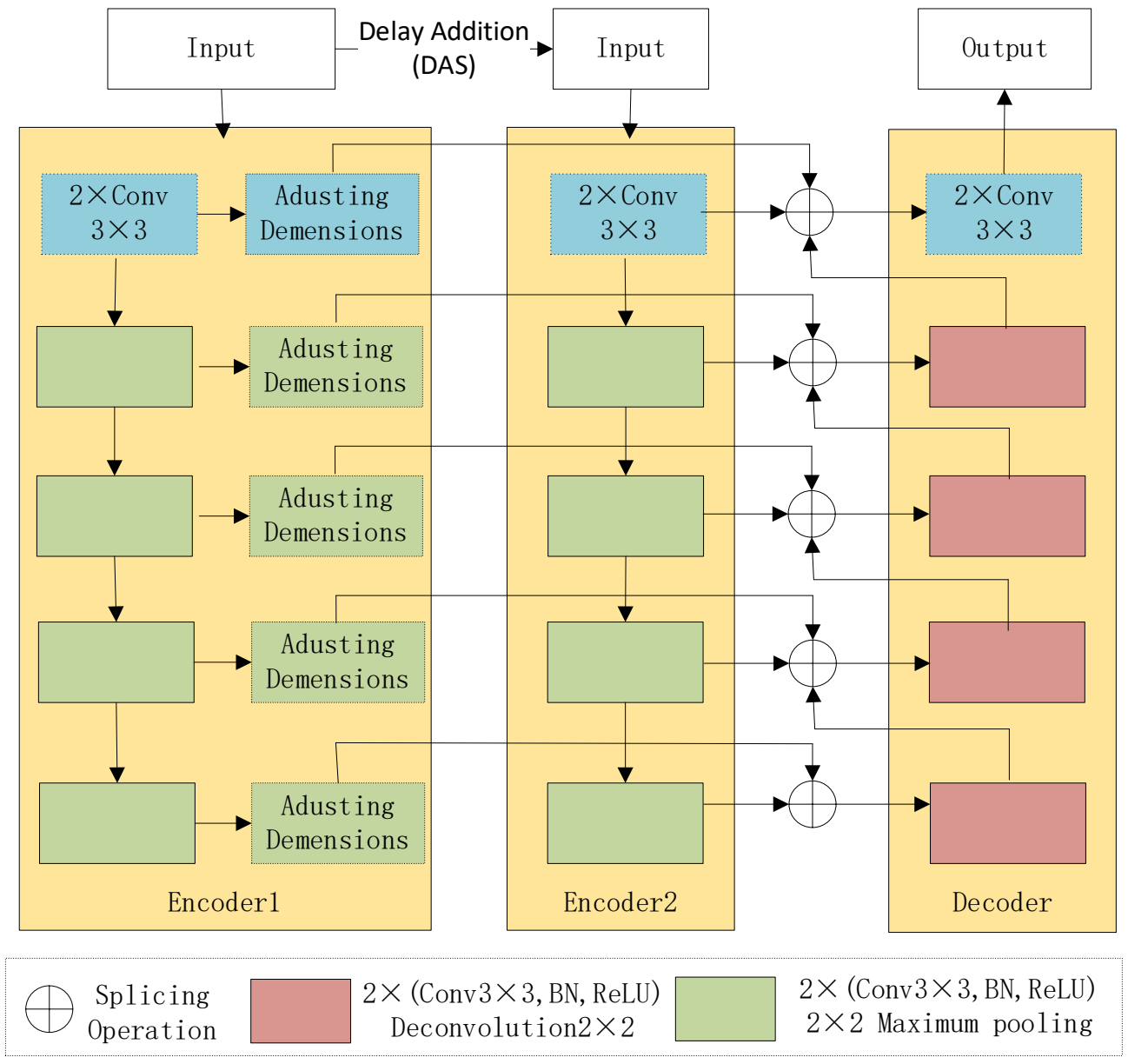

Figure 3. Y-Net Network Model (Lan et al.,2020)

The greatest difficulty in brain image segmentation lies in the low contrast of the surrounding tissues in the intracranial hemorrhage area. To enhance segmentation performance, Kuang et al. proposed Ψ-Net [16], as illustrated in Figure 4. The network is structured in the shape of the Greek letter "Ψ" and consists of three parallel encoders and a single decoder. The encoders process the target slice along with its two adjacent slices, incorporating self-attention mechanisms in the encoding layers and contextual attention mechanisms in the decoding layers. This design not only enhances the model's ability to capture global features and suppress superfluous information but also improves the restoration of local details through contextual attention mechanisms.

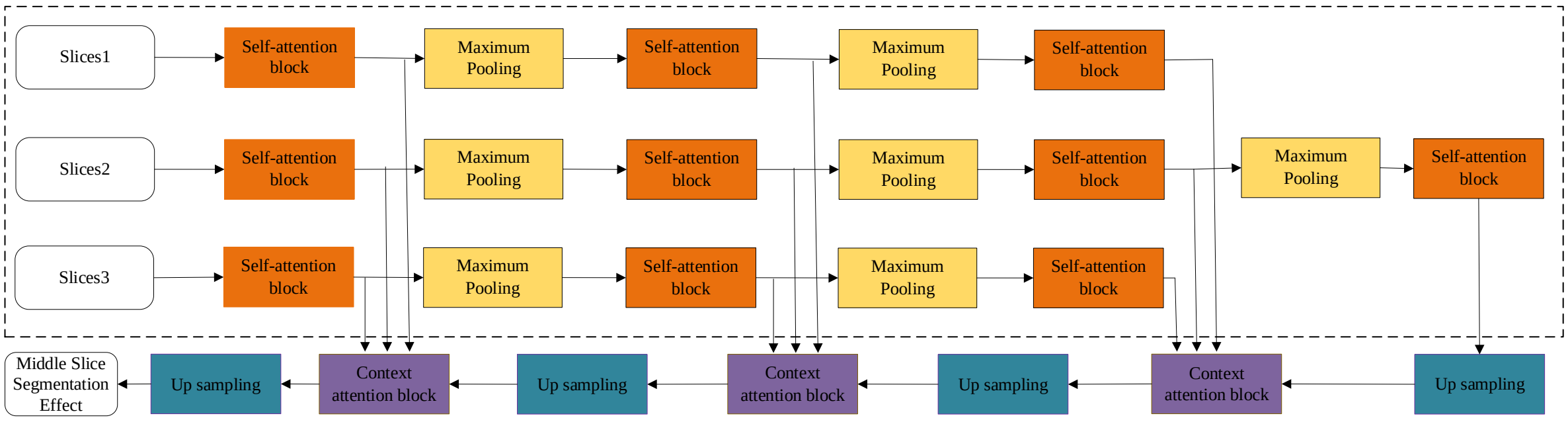

Figure 4. Ψ-Net Network Model( kuang et.al.,2020 )

Low Information Utilization of Multimodal Data in Ischemic Stroke Lesion Segmentatio n Methods. To address the challenge of low information utilization in ischemic stroke lesion s egmentation, Dolz et al. proposed a multipath dense U-Net model [17], which effectively extr acts features from diverse brain imaging modalities. The inputs to the encoding path consist of four modules, as illustrated in Figure 5. These multimodal inputs enhance the model's capacity to capture and integrate critical information, thereby improving segmentation performance.

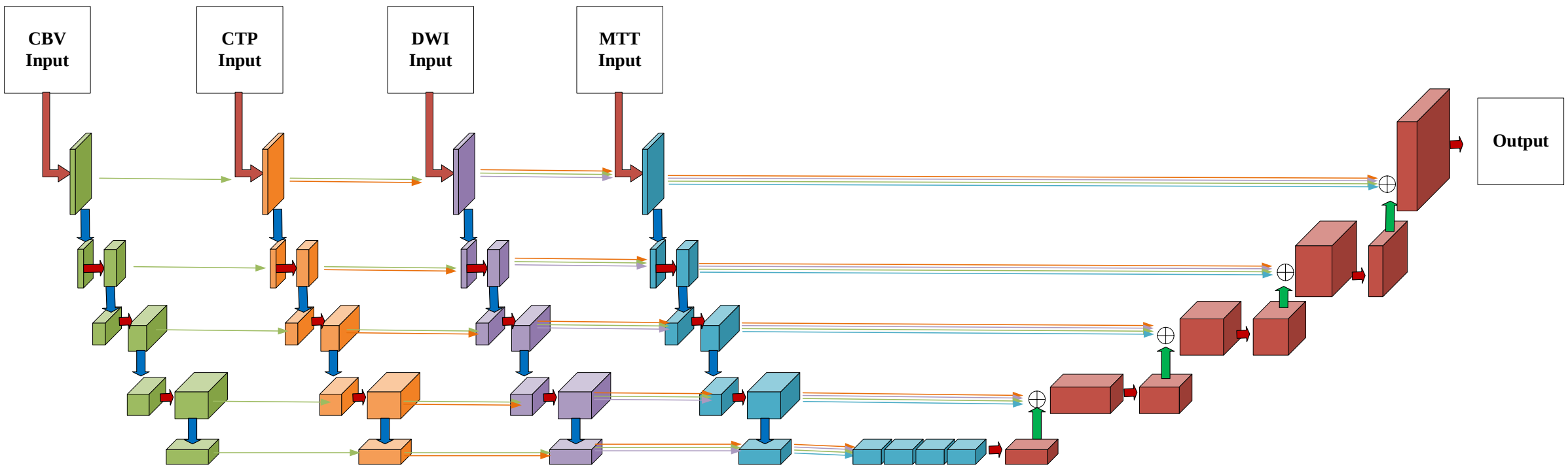

Figure 5. Multipath dense U-Net Model (Dolz et al.,2019)

## 2.2 Multiple U-Net networks

The networks use multiple U-Nets to obtain more features and multiscale information of images, thereby improving segmentation accuracy. Multiple U-Net concatenation takes the previous network's output as the subsequent network's input. Xia and Kulis proposed a completely unsupervised W-Net[18] (as shown in Figure 6). The U-Net network is nested into the U-shaped framework, with the left and right U-Nets representing the contraction and expansion paths, respectively. The network iteratively minimizes the reconstruction error of the decoder to include more input information in the encoding process, and the encoder's soft normalized pruning updates the gradient in backpropagation. Hu et al. proposed a coupled U-shaped architecture[19][20] that connects blocks with the same resolution to subsequent U-Net layers, resulting in a lightweight U-Net model. By reducing redundant parameters and improving parameter efficiency, this design ultimately enhances the stacked network's efficiency. Kang et al. introduced the CMUnet[21] network designed for solving the problem of sample scarcity by using modular and cascaded pretrained U-Nets for document image binarization. Cascaded U-Net reduces the problem of useful feature loss in each layer by using skip connections between modules in a serial manner.

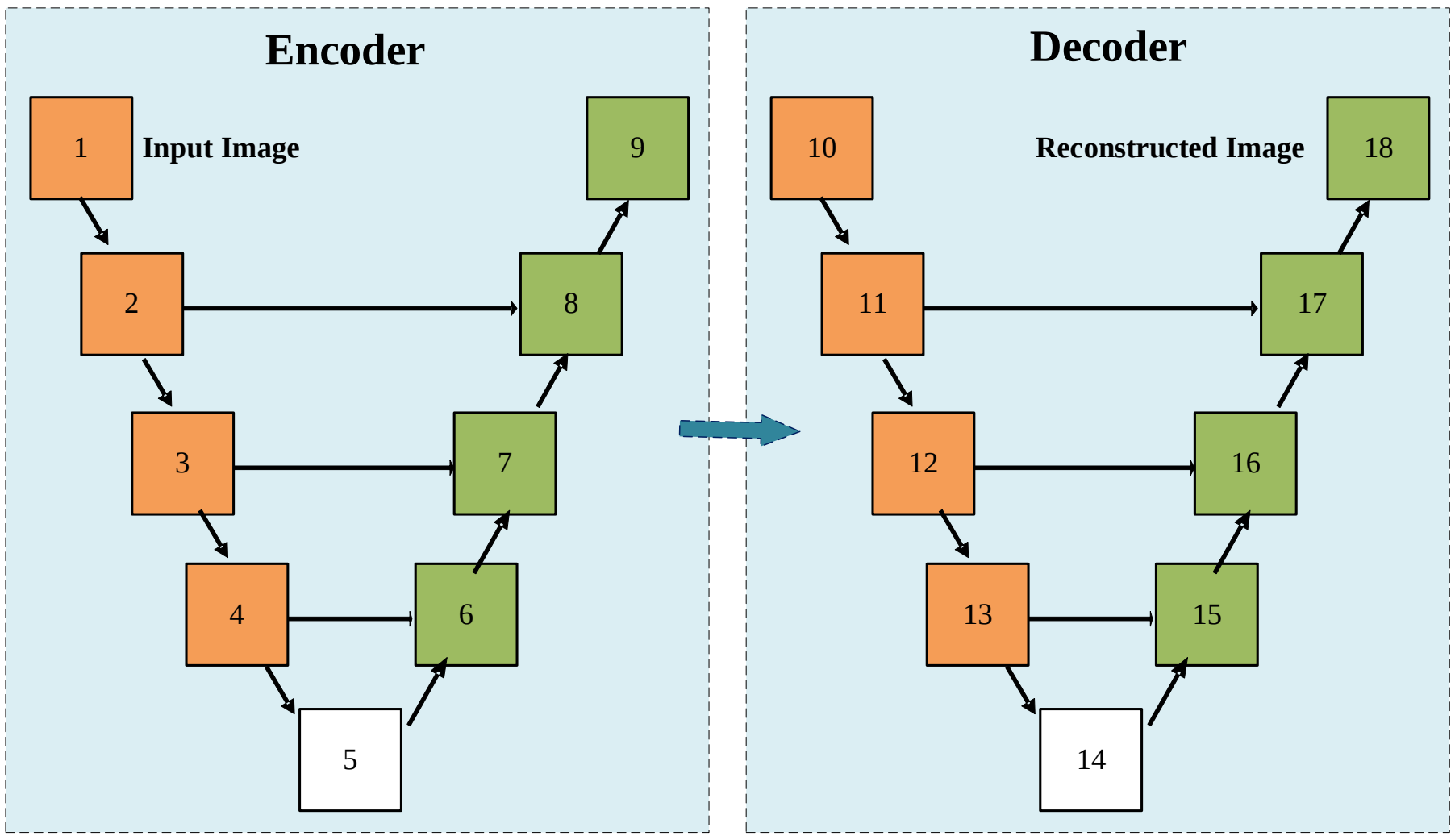

Figure 6. W-Net (Xia and Kulis,2017)

Multiple parallels combine U-shaped networks in a parallel manner, assign different functions to different U-Net networks, and input different types of images. Combining U-Net with other models. Zhao et al. proposed a deep learning model Triple U-net [22]. specifically designed for the segmentation of pathological cancer nuclei This model consists of three parallel branches: the RGB branch extracts color features from H&E-stained images, whereas the H branch focuses on capturing nuclear contour features from hematoxylin concentration images. The third branch, the segmentation branch, uses a module to fuse the features obtained from the RGB and H branches, thereby producing accurate nuclear segmentation results. By integrating multimodal information, Triple U-Net significantly enhances the accuracy and robustness of cancer nucleus segmentation in pathological images. Lee et al. proposed a network Mu-net [23] for denoising two two-photon microscope 3D images. The network consists of multiple parallel U-shaped networks with different scales, where processes of images with different scales are proposed by different networks. The network continuously downsamples the image to generate the input subsequent U-Net network, achieving image reconstruction from coarse to fine and constructing low- to high-frequency target outputs. Li et al. proposed a supervised U-shaped method [24] that incorporates supervision at the bottleneck layer, where the most condensed features are located. This supervision guides the model to extract more discriminative and correlated features during the encoding process, which enhances the segmentation performance. By applying additional supervision at this critical layer, the model can better capture complex patterns and improve the accuracy and robustness of the segmentation.

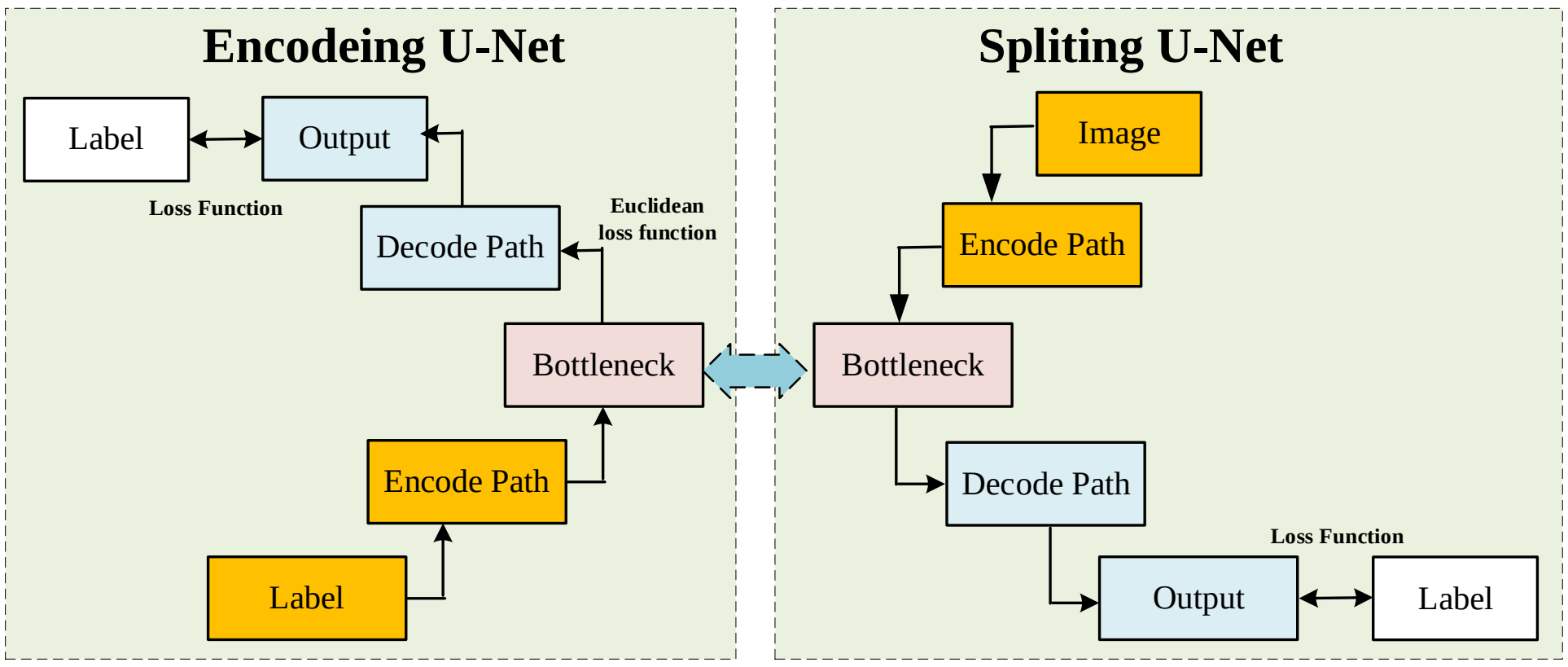

Figure 7. Bottleneck feature supervised U-Net (Li et al. ,2020)

## 2.3 3D U-Net Networks

Compared with 2D images, 3D medical images contain richer spatial structural information, providing a more comprehensive three-dimensional view of organs and lesions. This enhanced perspective can aid in more accurate diagnosis and treatment planning by medical professionals. Owing to its architectural design, 3D U-Net can directly process three-dimensional data, fully utilize spatial information in images and capture feature relationships across multiple dimensions. This enables the model to achieve higher accuracy and stability in complex medical image segmentation tasks. However, despite the significant advantages of 3D U-Net in handling three-dimensional medical images, it also faces several challenges. For example, training a 3D U-Net model, particularly with high-resolution volumetric images, typically requires substantial memory (VRAM) and computational resources.

Cicek et al. designed a 3D U-Net [2] network model for segmenting three-dimensional medical images. The model employs 3D convolutional kernels to process volumetric data while maintaining the "U-shaped" architecture. It consists of a contracting path and an expansive path, with skip connections between the corresponding layers to transfer multiscale feature information. This design effectively leverages spatial structure information, enhancing the model's ability to understand three-dimensional structures. By simultaneously capturing high-resolution detail features and contextual information, the model improves segmentation accuracy. Figure 8 shows the details of the 3D U-Net architecture, and the primary structural layout is detailed in Table 2.

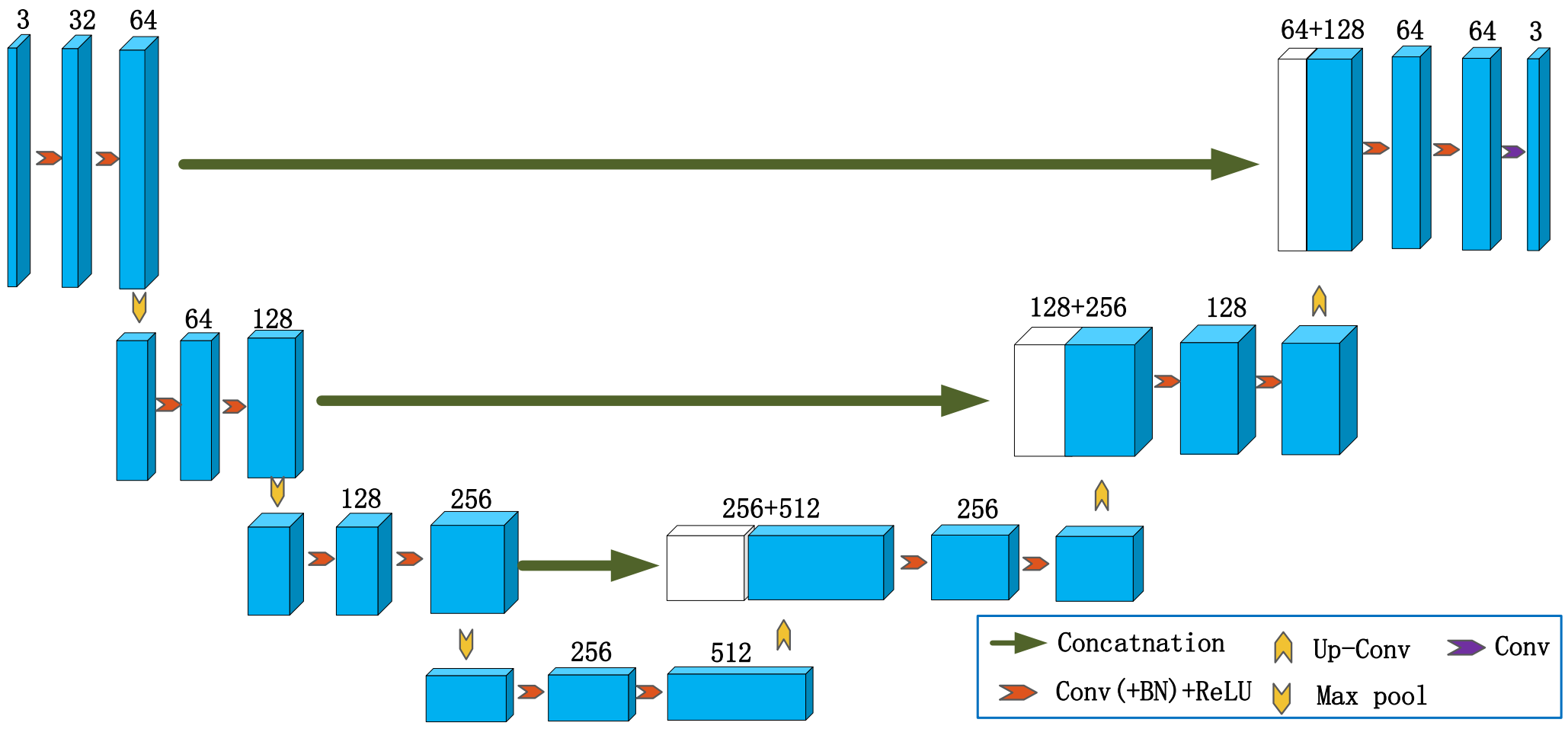

Figure 8. The 3D U-Net model architecture (Çiçek et al.,2016)

Table 2 3D U-Net Encoder and Decoder Details

| Layer Type | Input Shape | Output Shape | Description |
|---|---|---|---|
| Input | 128×128×128×3 | 128×128×128×3 | Input volume |
| Conv3D Block 1 | 128×128×128×3 | 128×128×128×8 | Conv3D × 2 |
| MaxPooling3D | 128×128×128×8 | 64×64×64×8 | Downsampling |
| Conv3D Block 2 | 64×64×64×8 | 64×64×64×16 | Conv3D × 2 |
| MaxPooling3D | 64×64×64×16 | 32×32×32×16 | Downsampling |
| Conv3D Block 3 | 32×32×32×16 | 32×32×32×32 | Conv3D × 2 |
| MaxPooling3D | 32×32×32×32 | 16×16×16×32 | Downsampling |
| Conv3D Block 4 | 16×16×16×32 | 16×16×16×64 | Conv3D × 2 |
| MaxPooling3D | 16×16×16×64 | 8×8×8×64 | Downsampling |
| Conv3D Block 5 | 8×8×8×64 | 8×8×8×128 | Conv3D × 2 |
| UpSampling Block 1 | 8×8×8×128 | 16×16×16×64 | UpSampling + Skip + Conv3D |
| UpSampling Block 2 | 16×16×16×64 | 32×32×32×32 | UpSampling + Skip + Conv3D |
| UpSampling Block 3 | 32×32×32×32 | 64×64×64×16 | UpSampling + Skip + Conv3D |
| UpSampling Block 4 | 64×64×64×16 | 128×128×128×8 | UpSampling + Skip + Conv3D |
| Output | 128×128×128×8 | 128×128×128×4 | Final output volume |

Milletari et al. introduced a new form of U-Net architecture specifically designed for 3D medical image segmentation tasks. V-Net [3] retains the fundamental "U-shaped" network structure but is optimized for processing 3D volumetric data. As illustrated in Figure 9, V-Net incorporates several key modifications to the original U-Net architecture, including the use of 3D convolutional layers and 3D pooling operations. Additionally, the network integrates convolutional residual units to enhance feature extraction across multiple spatial scales. These improvements enable V-Net to capture three-dimensional spatial dependencies, making it suitable for three-dimensional object segmentation tasks.

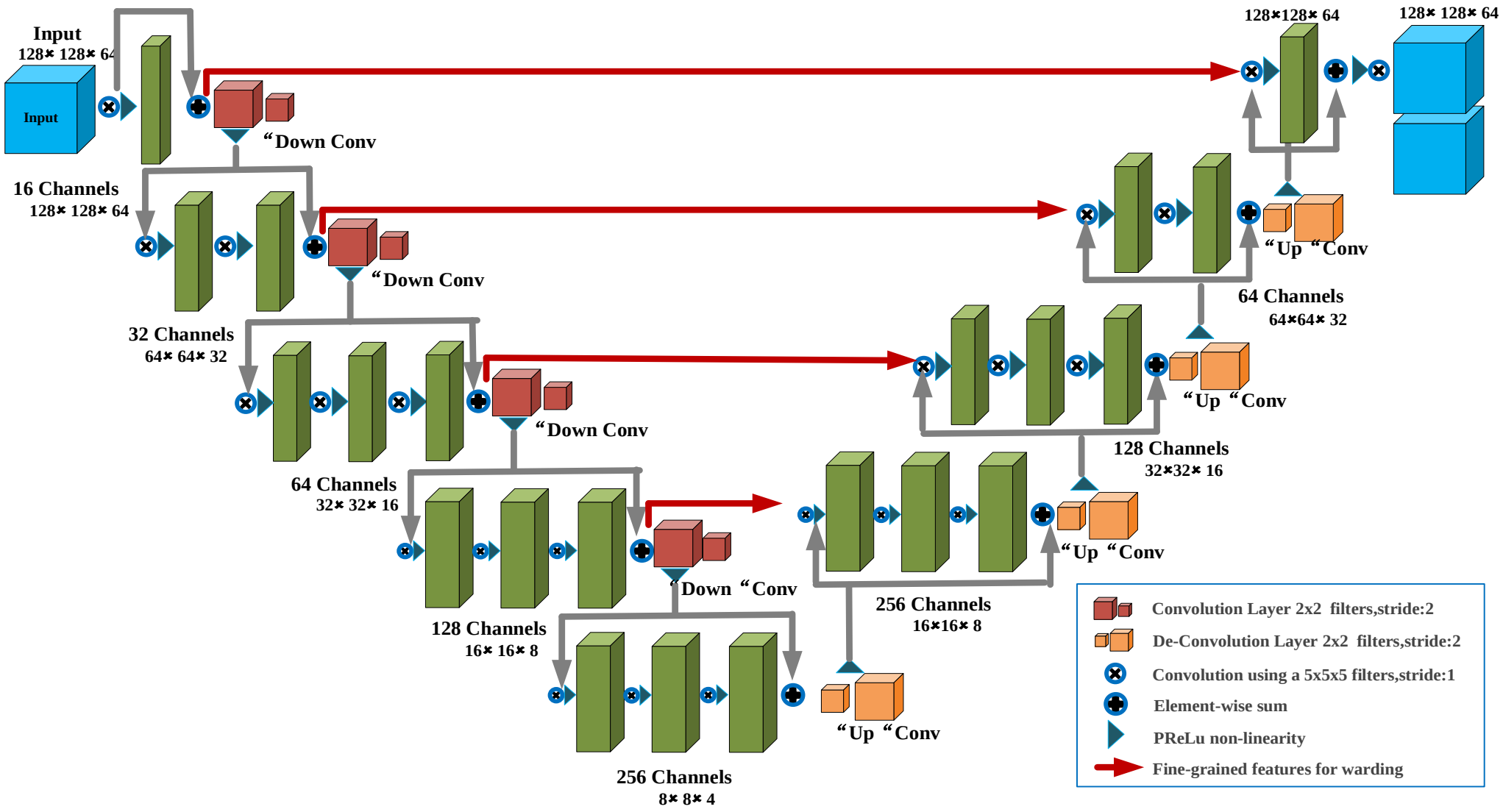

Figure 9. The structure of the V-Net architecture (Milletari F et al.,2016)

Unlike U-Net, V-Net uses 3D convolutional layers, allowing for feature extraction in three-dimensional space and capturing spatial information in 3D images. The encoding path of V-Net incorporates residual units, which add the input to the output via skip connections directly. The method not only alleviates the vanishing gradient but also enhances the efficiency of feature extraction. The encoder and decoder details of V-Net are shown in Table 3.

Table 3 Encoder and Decoder details of V-Net

| Layer | Input Shape | Output Shape | Description |
|---|---|---|---|
| Input | 128×128×128×3 | 128×128×128×3 | Input image |
| Conv3D Block 1 | 128×128×128×3 | 128×128×128×8 | Conv3D × 2, Stride 1 |
| MaxPooling3D | 128×128×128×8 | 64×64×64×8 | Stride 2 |
| Conv3D Block 2 | 64×64×64×8 | 64×64×64×16 | Conv3D × 2, Stride 1 |
| MaxPooling3D | 64×64×64×16 | 32×32×32×16 | Stride 2 |
| Conv3D Block 3 | 32×32×32×16 | 32×32×32×32 | Conv3D × 2, Stride 1 |
| MaxPooling3D | 32×32×32×32 | 16×16×16×32 | Stride 2 |
| Conv3D Block 4 | 16×16×16×32 | 16×16×16×64 | Conv3D × 2, Stride 1 |
| MaxPooling3D | 16×16×16×64 | 8×8×8×64 | Stride 2 |
| Conv3D Block 5 | 8×8×8×64 | 8×8×8×128 | Conv3D × 2, Stride 1 |
| Deconv Block 1 | 8×8×8×128 | 16×16×16×64 | Deconv + Skip Connection |
| Conv3D Block 6 | 16×16×16×128 | 16×16×16×64 | Conv3D × 2, Stride 1 |
| Deconv Block 2 | 16×16×16×64 | 32×32×32×32 | Deconv + Skip Connection |
| Conv3D Block 7 | 32×32×32×64 | 32×32×32×32 | Conv3D × 2, Stride 1 |
| Deconv Block 3 | 32×32×32×32 | 64×64×64×16 | Deconv + Skip Connection |
| Conv3D Block 8 | 64×64×64×32 | 64×64×64×16 | Conv3D × 2, Stride 1 |
| Deconv Block 4 | 64×64×64×16 | 128×128×128×8 | Deconv + Skip Connection |
| Output | 128×128×128×8 | 128×128×128×4 | Final segmentation map |

Hatamizadeh et al. proposed a new “U-shaped” architecture UNETR [25], the details of which are shown in Figure 10. This architecture combines convolutional neural networks (CNNs) with self-attention mechanisms, enabling efficient feature extraction and global context integration. Multiresolution skip connections serve as bridges, connecting transformer encoders and decoders, allowing for accurate segmentation of complex medical images.

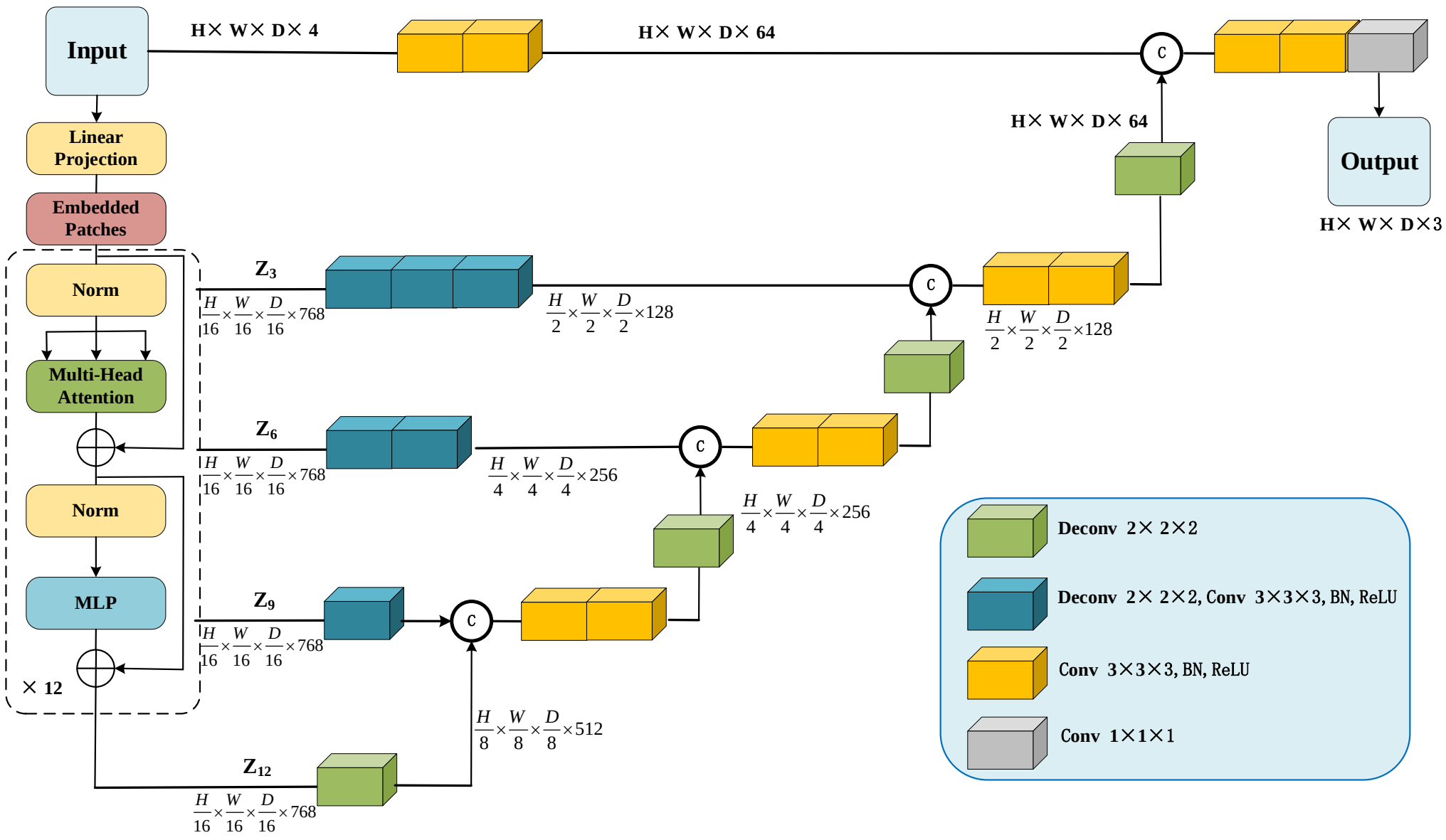

Figure 10. UNETR Architecture (Hatamizadeh et al. 2022)

Aboussaleh et al. designed a hybrid network model using multimodal magnetic resonance imaging, which performs well in brain tumor segmentation. This model combines the strengths of 3D U-Net and ResNet, integrating 3D convolutions and residual blocks within the encoder path. By incorporating multimodal MRI data and utilizing skip connections, the model effectively preserves high-resolution spatial details while simultaneously enhancing the fusion of different modal features. Details of the network structure of 3D U-VNet [9] are shown in Figure 11.

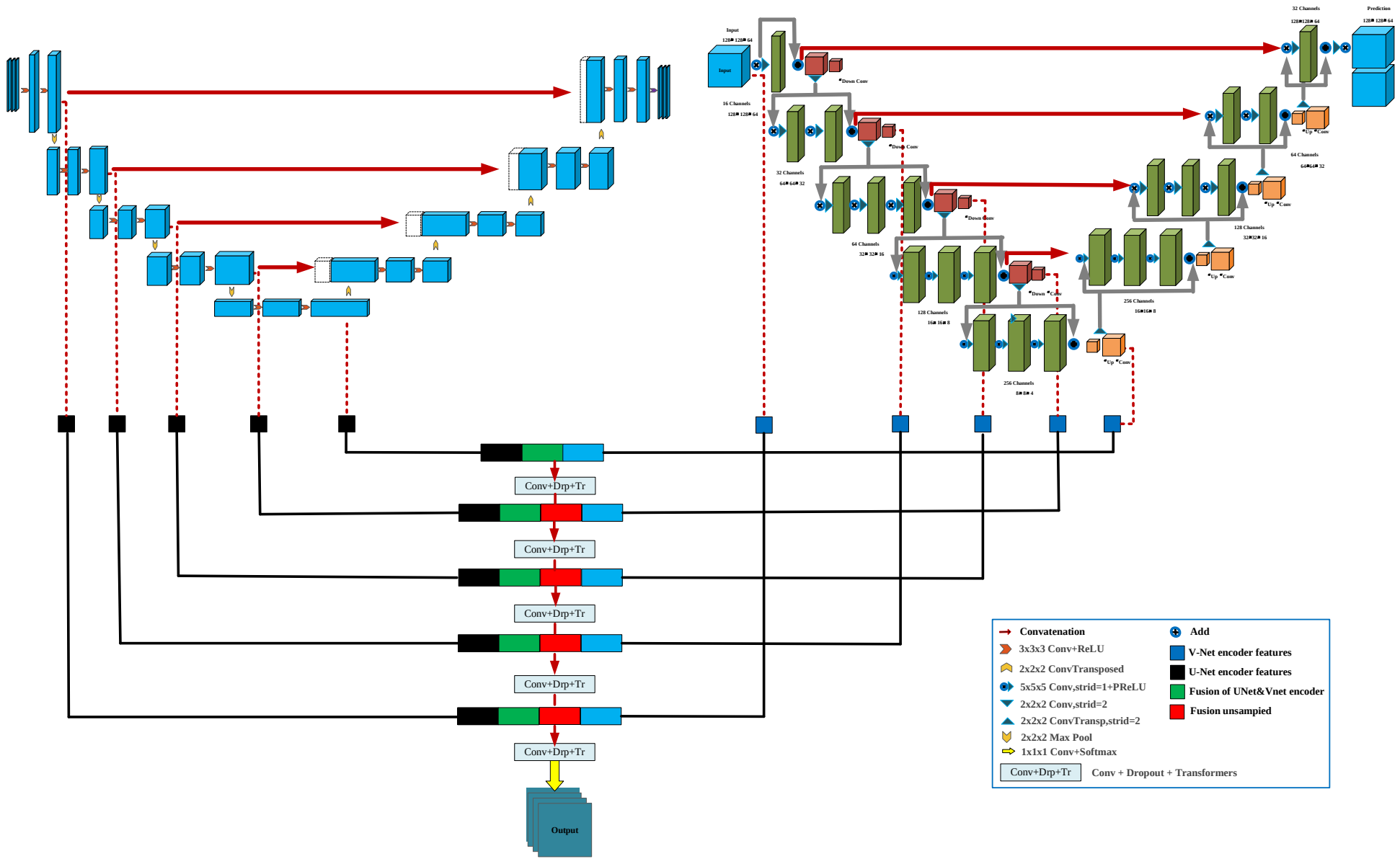

Figure 11. Overview of the 3DUV-NetR+ architecture (Aboussaleh et al. 2024).

Dou et al. designed a three-dimensional supervised network [26] for liver CT image segmentation and heart and vascular MR image segmentation. This network introduces a multilevel supervision mechanism, enabling supervised learning at different layers of the network. Segmentation predictions are generated at each level and progressively integrated, thereby enhancing the training efficiency and segmentation accuracy while ensuring stable performance. Chen et al. proposed the VoxResNet [27] network, which can fully utilize the interlayer information of 3D images to segment brain tissue images, addressing the issue of decreased segmentation performance due to deeper network depth, and use residual blocks to deepen voxel blocks to obtain more multimodal and contextual information, alleviating network degradation. Some researchers conceptualize the task of 3D medical image segmentation as a prediction task between sequences.

### 2.4 Semi-supervised and unspuervised Domain Adaptation

In the field of medical image segmentation, the scarcity of annotated data remains a common challenge. Due to the limited availability of annotated datasets, constructing pseudo-labels with strong intra-class compactness and inter-class separability proves to be a significant difficulty, which in turn constrains the improvement of model performance. Consistency learning has emerged as a promising approach to address this issue by effectively leveraging the limited annotated data while integrating abundant unlabeled data. However, two critical challenges persist in its practical application: enhancing prediction diversity and ensuring training stability. Addressing these challenges is essential to further improve the segmentation performance and robustness of such models.

Zhenxi Zhang .et, al. introduces the **Self-aware and Cross-sample Prototype Network (SCP-Net)**[28], which generates two distinct types of prototype predictions to enhance semantic information interaction and ensure controlled inconsistency during consistency training. Additionally, the model leverages the predictive uncertainty between the self-aware prototype prediction and multiple prediction results to reweight the consistency constraint loss for cross-sample prototypes. This approach mitigates the adverse effects of label noise in challenging regions, such as low-contrast areas or adhesive boundaries, thereby facilitating a more stable consistency training process and enhancing the model's performance and accuracy. Details of the architecture of SCP-Net are shown in Figure 12.

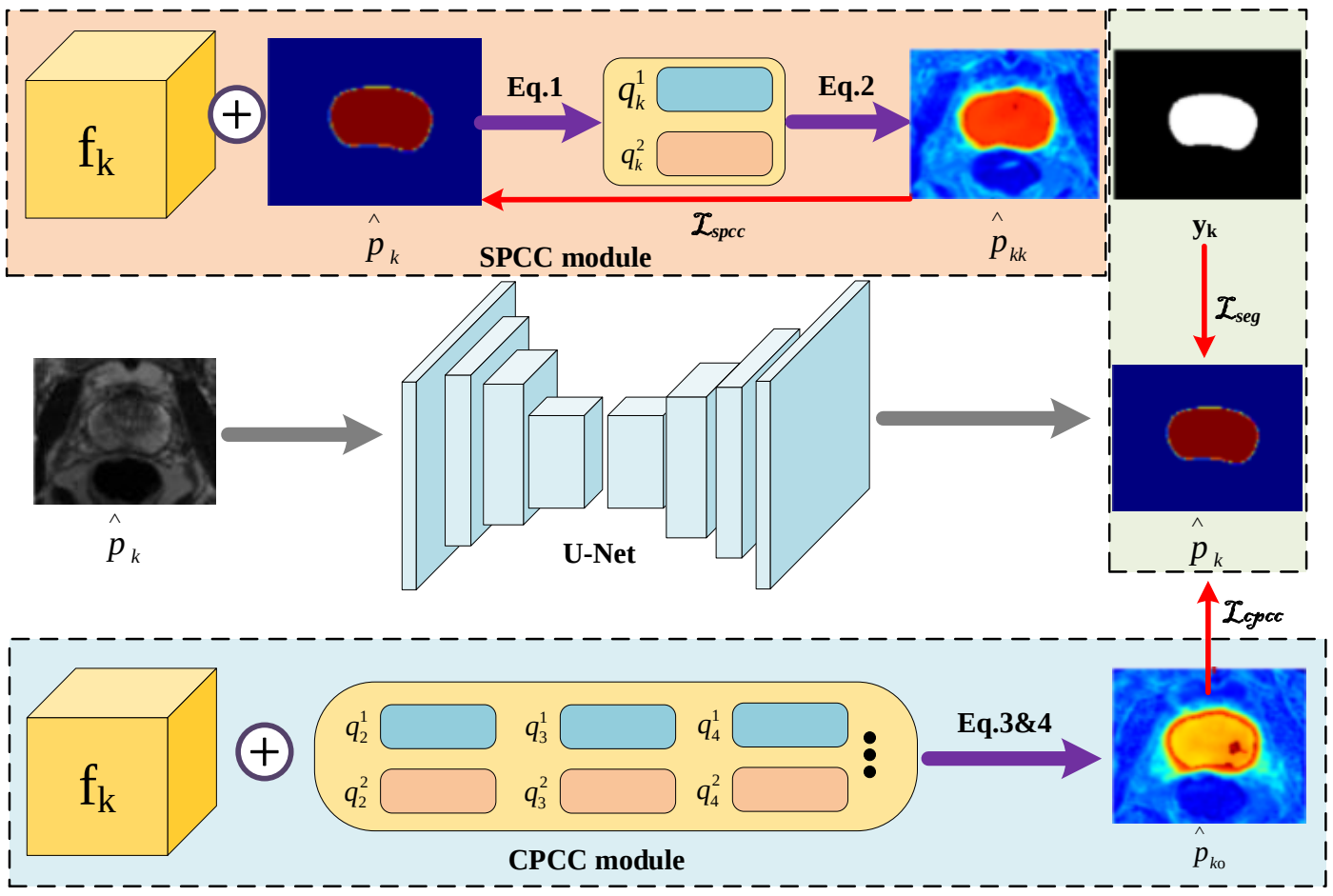

Figure 12 Detail of the architecture of SCP-Net (Zhenxi Zhang et al.,2023)

Torbunov Dmitrii et.al. introduces UVCGAN, a CycleGAN variant that integrates a UNet generator with a Vision Transformer (ViT) bottleneck, augmented by advanced training techniques such as gradient penalty and self-supervised pre-training. The model approach addresses critical challenges in unpaired image-to-image translation, particularly in applications requiring stringent cycle-consistency and high correlation between input and output images. The UVCGAN framework demonstrates significant improvements over existing models on multiple benchmark datasets, including Selfie2Anime and CelebA-derived tasks (Gender Swap and Eyeglasses). The model achieves superior results in Fréchet Inception Distance (FID) and Kernel Inception Distance (KID) metrics, underscoring its capability to preserve essential features while ensuring high visual fidelity. The incorporation of the ViT bottleneck enhances the generator's ability to learn non-local dependencies, thereby improving its capacity to capture complex spatial patterns. Details of the network structure of UVCGAN are shown in Figure 13.

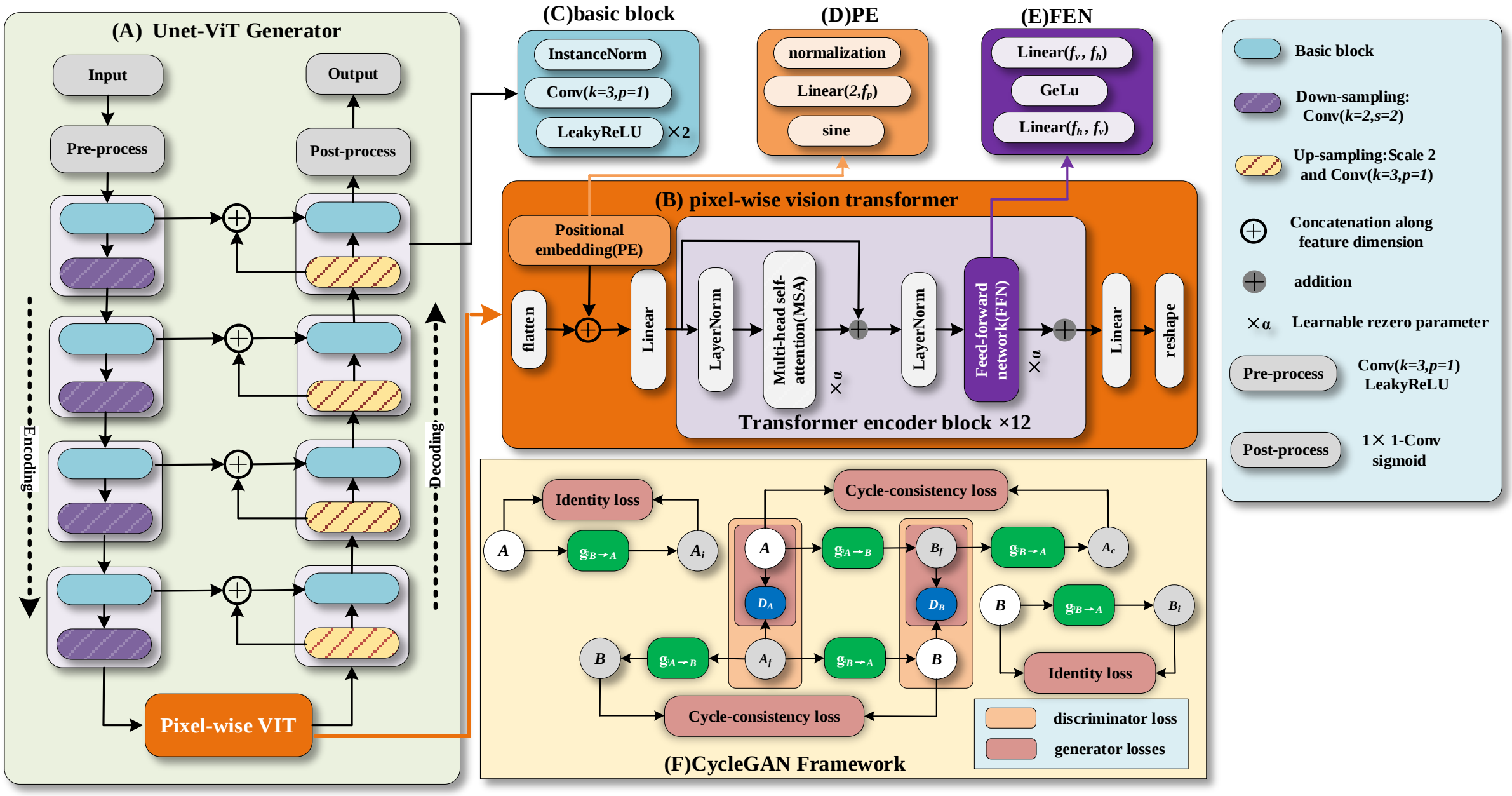

Figure 13 The architecture of UVCGAN ( Torbunov Dmitrii et.al.,2022)

Rigorous evaluation through standardized testing protocols and comprehensive ablation studies underscores the pivotal roles of **gradient penalty** and **self-supervised pretraining** in achieving optimal performance. However, the inherent complexity of the model's architecture and its reliance on extensive hyperparameter tuning limit its generalizability and practical applicability. Future research should prioritize enhancing computational efficiency and reducing the dependence on manual hyperparameter adjustments to improve the model's scalability and adaptability across diverse settings.

In summary, this study makes a noteworthy contribution to the field of **unpaired image-to-image translation**. By seamlessly integrating a transformer-based architecture into the CycleGAN framework, the authors have established a strong foundation for advancing more robust, versatile, and high-performing models capable of addressing a wide range of challenging translation tasks.

Boyun Zheng et al proposes a novel framework - Semantic Preserved Dual Domain Distribution Interrupt (DDSP)[30], to solve the domain transformation problem in medical image segmentation tasks. The traditional unsupervised domain adaptation (UDA) method based on GAN suffers from instability and semantic inconsistency due to incomplete domain transformation, as shown in Figure 14.

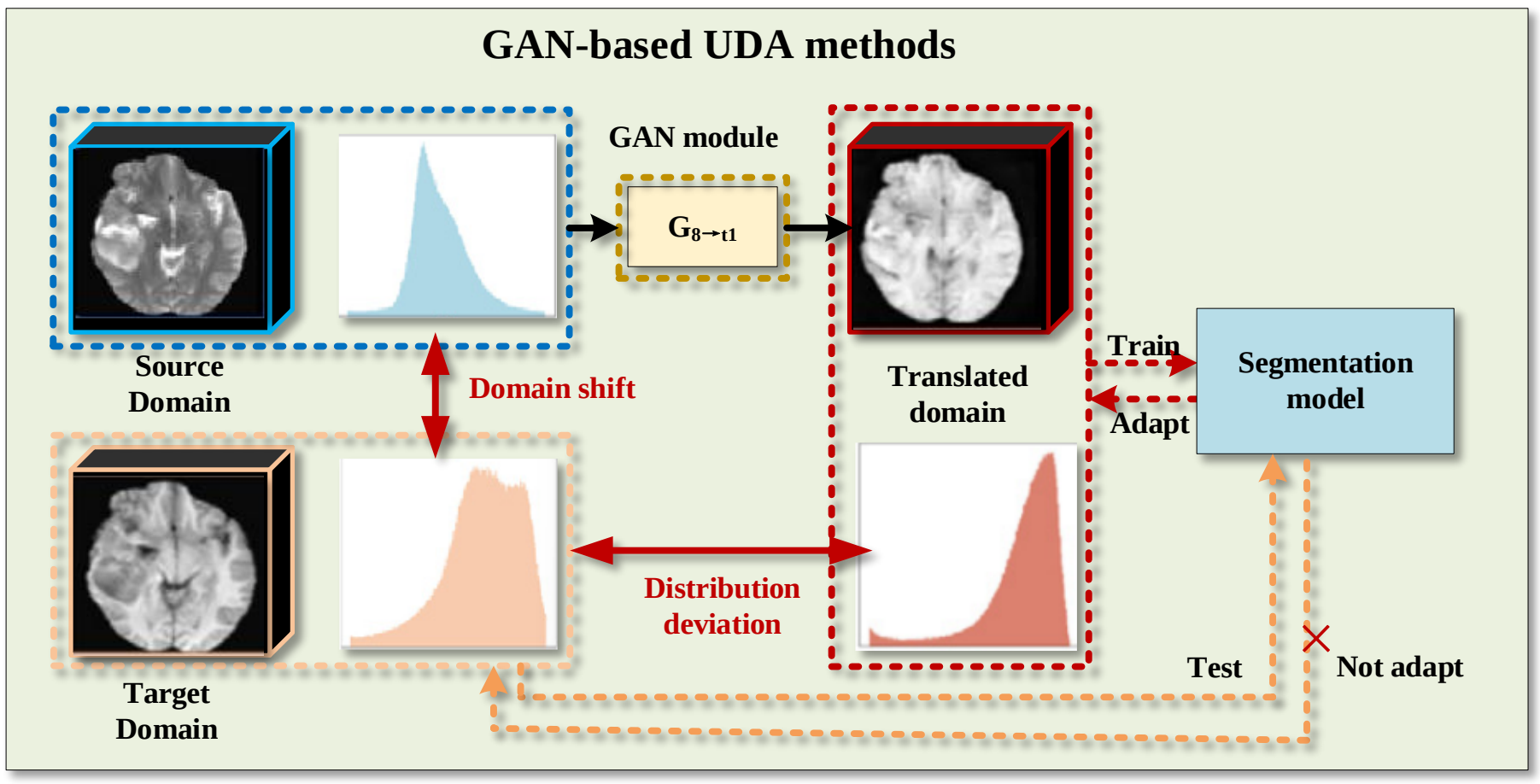

Figure 14 The GAN based on UDA methods.

The DDSP framework introduces key innovations, notably the inter-channel similarity feature alignment (IFA) and semantic consistency loss, which facilitate precise alignment of source and target domain features while preserving semantic integrity. Rigorous validation on three public datasets—cardiac, brain tumor, and prostate segmentation—demonstrates the framework's superior performance over state-of-the-art UDA methods. Remarkably, DDSP achieves Dice similarity scores that approach those of fully-supervised models, underscoring its efficacy and potential impact. Details of the architecture of DDSP are shown in Figure 15.

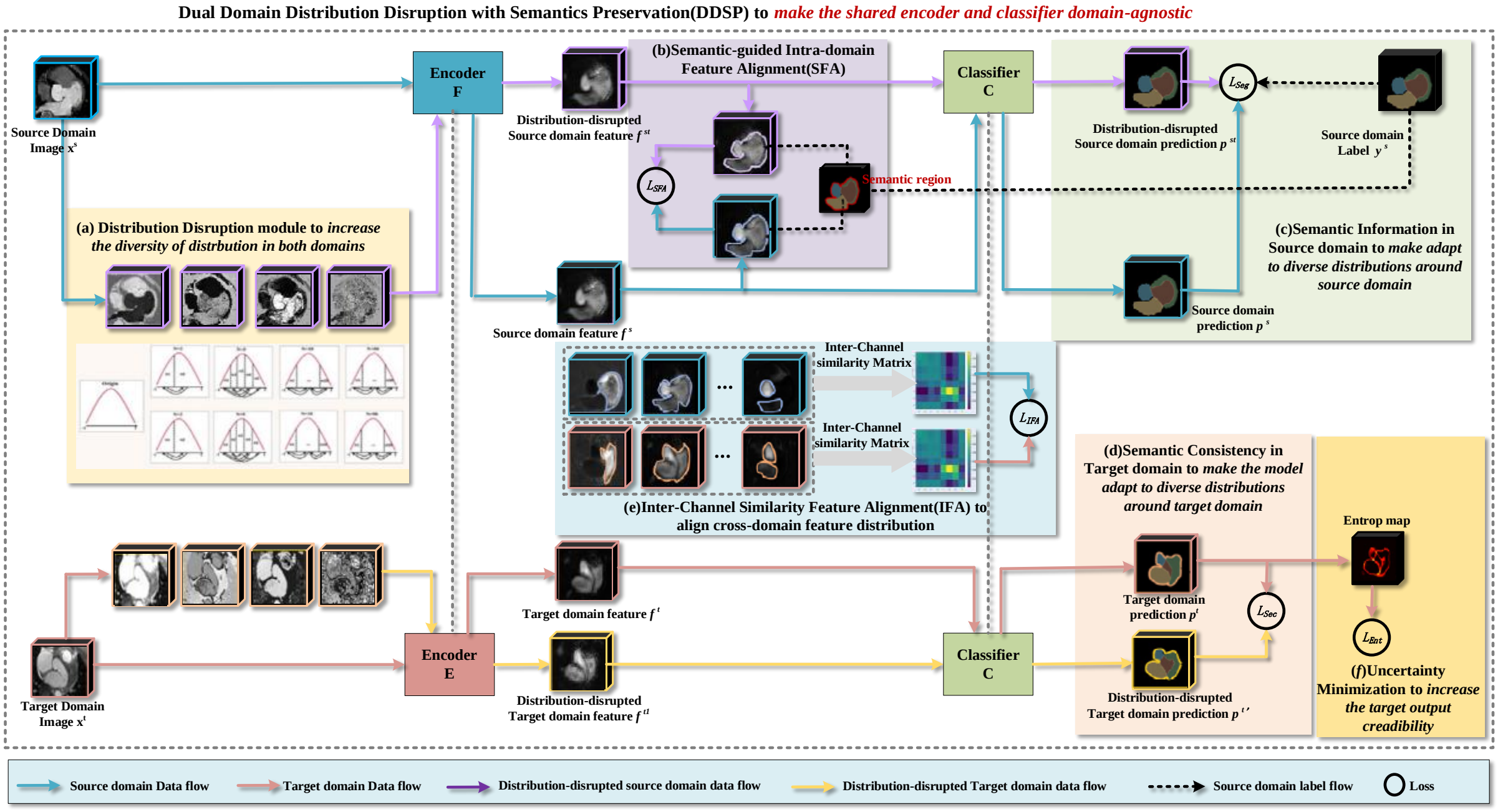

Figure 15 The architecture of DDSP framework (Boyun Zheng et al.,2024)

However, the reliance on **manual parameter tuning** for distribution transformations and the limited exploration of **computational efficiency** may restrict its scalability and applicability in diverse clinical scenarios. Future research should focus on automating parameter optimization and extending validation to encompass additional imaging modalities and datasets to enhance the framework's robustness and practicality.

In summary, this study offers a significant contribution to medical image segmentation by addressing fundamental limitations of GAN-based UDA methods, including instability and semantic inconsistencies. By providing a robust and domain-agnostic approach, the DDSP framework paves the way for more reliable and scalable solutions, advancing the clinical utility of segmentation models in diverse medical imaging applications.

## 3. Improvement methods and strategies

To improve the original U-Net model in terms of network performance, efficiency, accuracy, and generalizability, several enhancements have been introduced. These improvements encompass various aspects of the U-Net architecture, including data augmentation, convolution operations, downsampling operations and upsampling operations, model optimization strategies, and skip connections. These modifications collectively contribute to a more robust and effective U-Net model capable of superior performance across diverse applications.

### 3.1 Data Enhancement.

Data augmentation plays a crucial role in enhancing U-Net models by artificially expanding the training dataset, which helps mitigate overfitting and improves the model's generalizability. Data augmentation creates a more diverse and extensive set of training samples by applying various transformations, such as rotations, translations, scaling, and flipping, to the original images. This diversity allows the U-Net model to learn more robust features, making it better equipped to handle variations and noise in unseen data. Additionally, data augmentation can simulate different imaging conditions and scenarios, further strengthening the model's ability to perform accurately and reliably in real-world applications. Data augmentation techniques ensure that the U-Net model is trained on a comprehensive dataset, leading to enhanced performance, increased accuracy, and improved efficiency in segmentation tasks.

Methods of data enhancement can be roughly divided into two categories: basic data augmentation and advanced data augmentation. Basic data augmentation primarily involves modifications to the position or color of the image. In contrast, advanced data augmentation mainly refers to image blending and automatic augmentation. The specific categorization information is shown in Figure 16.

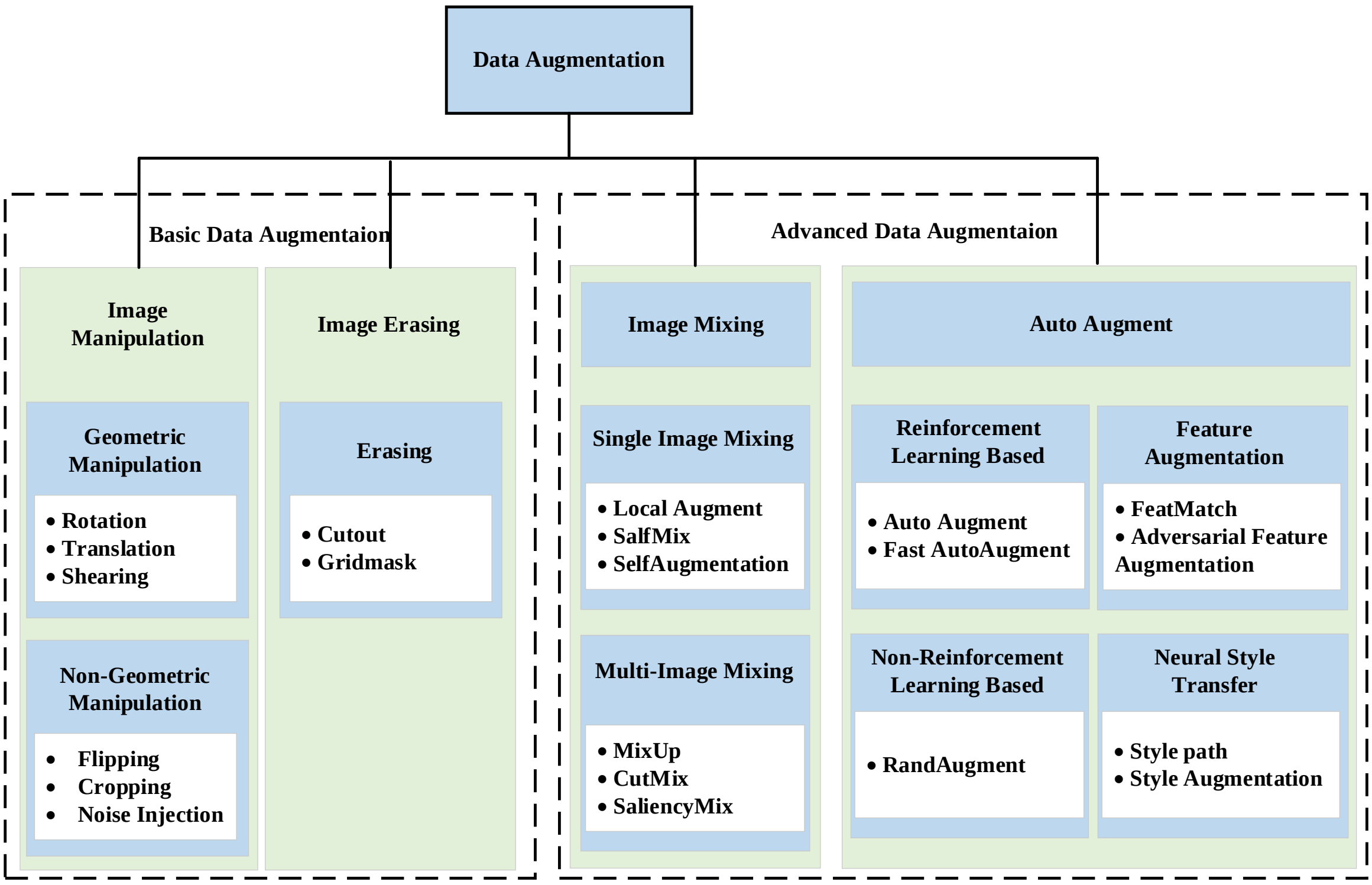

Figure 16. Content of Data Augmentation

## 3.2 Network improvement mechanism.

An increase in the U-Net network's segmentation performance was achieved through residual connections, dense layers, attention mechanisms, and their combination.

### 3.2.1 Residual Neural Network Mechanism

ResNet [31] is a deep learning model proposed by He et al. In that year, it achieved first place in the ImageNet competition for both classification and object detection tasks. ResNet's main innovation lies in solving the degradation problem in deep neural network training, enabling the network to be effectively trained to very deep levels, breaking through the previous limitations of network depth. Ibtehaz and Rahman proposed a MultiResUNet [32] network for dermatoscopic image segmentation by replacing the convolution operation of shrinking and expanding paths with a MultiRes block. The MultiRes block obtains outputs from three convolutional blocks of sizes 3×3, 5×5, and 7×7 and obtains intermediate features of different scales through parallel concatenation. Khanna et al. proposed a residual U-Net [33] model for lung CT image segmentation. This model integrates residual blocks into the contracting path of the U-Net architecture, effectively preventing network degradation and significantly reducing computational consumption. Sarica et al. introduced a novel dense residual U-Net [34], a deep learning architecture aimed at improving image segmentation tasks. This model integrates dense connections with residual blocks within the U-Net framework, enabling more efficient feature propagation and reuse throughout the network. The combination of dense and residual connections greatly enhances the model's ability to capture features while maintaining a relatively low parameter count, thereby enhancing its ability to address complex segmentation tasks with greater accuracy and reduced computational cost. Its architecture is shown in Figure 17.

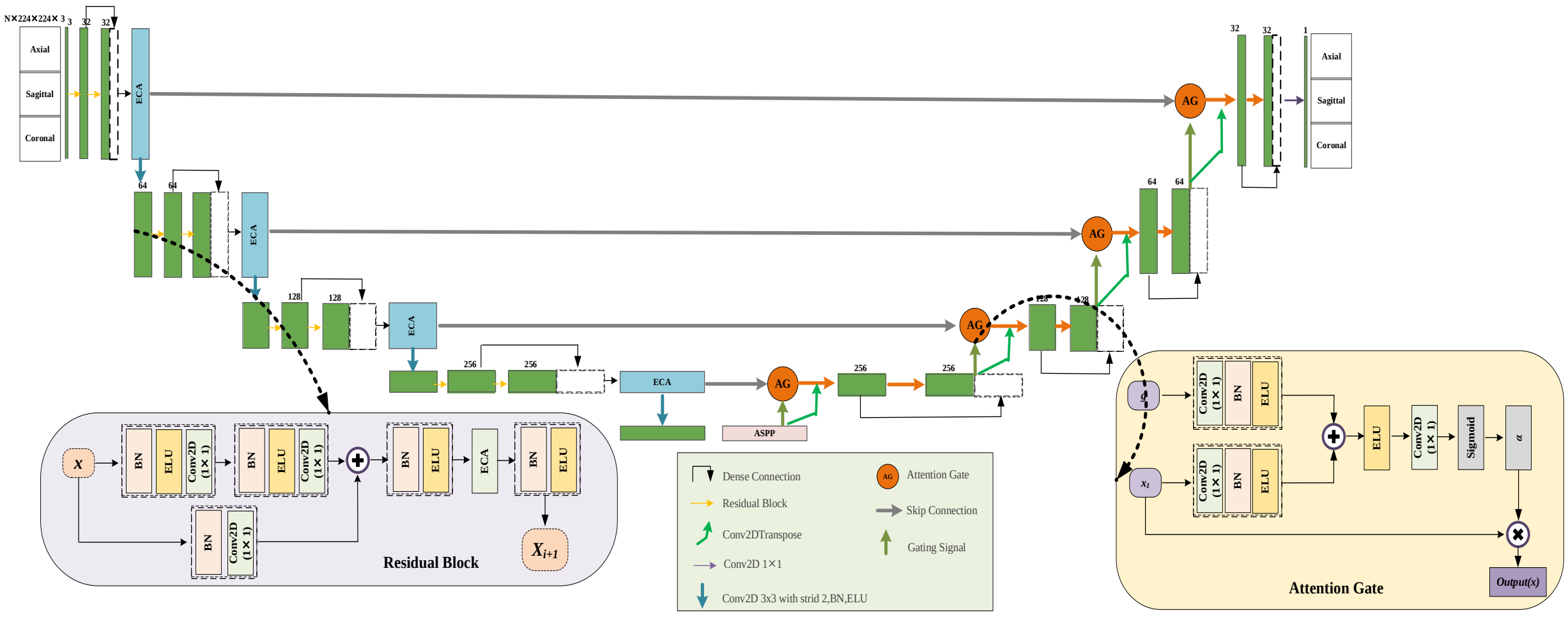

Figure 17. Residual U-Net architecture (Sarica et al.,2023)

### 3.2.2 Dense Convolutional Network Mechanism.

Huang et al. introduced a dense convolutional network based on ResNet [13], which establishes a parallel connection between the output and input. This effectively alleviates gradient vanishing, increases feature reusability, and minimizes the parameters used in network computation. Huang et al. proposed a novel UNet3+ [35] network that utilizes complete skip connections with extensive deep supervision. The method integrates detailed information merged with high-level semantic concepts across different scales, making it particularly effective for multiscale organ segmentation. The model approach improves accuracy while reducing network parameters, enhancing computational efficiency. Additionally, the structure of a combined loss function with a classification driver module is added to the network model to sharpen the boundaries, avoiding over segmentation and obtaining more accurate results. Xiang et al. proposed a bidirectional O-type network (BiO Net) [36] with skip connections added to the decoder path, establishing an O-shaped loop path between the encoder and decoder. This O-shaped model can be repeated recursively many times to improve performance without introducing additional parameters, which helps to avoid overfitting. Ibtehaz and Rahman proposed MultiResUNet [32], which uses a convolutional layer sequence to form residual skip connections, enhancing the network's learning ability and making it easier to delineate fuzzy boundaries. Liu et al. designed a segmentation model, TransUNet+ [37], with a feature enhancement module by redesigning skip connections. This network constructs the relationship between image features through the utilization of column vectors from a score matrix to strengthen feature representation. Figure 18 illustrates four different skip connection structures grounded in the U-Net framework.

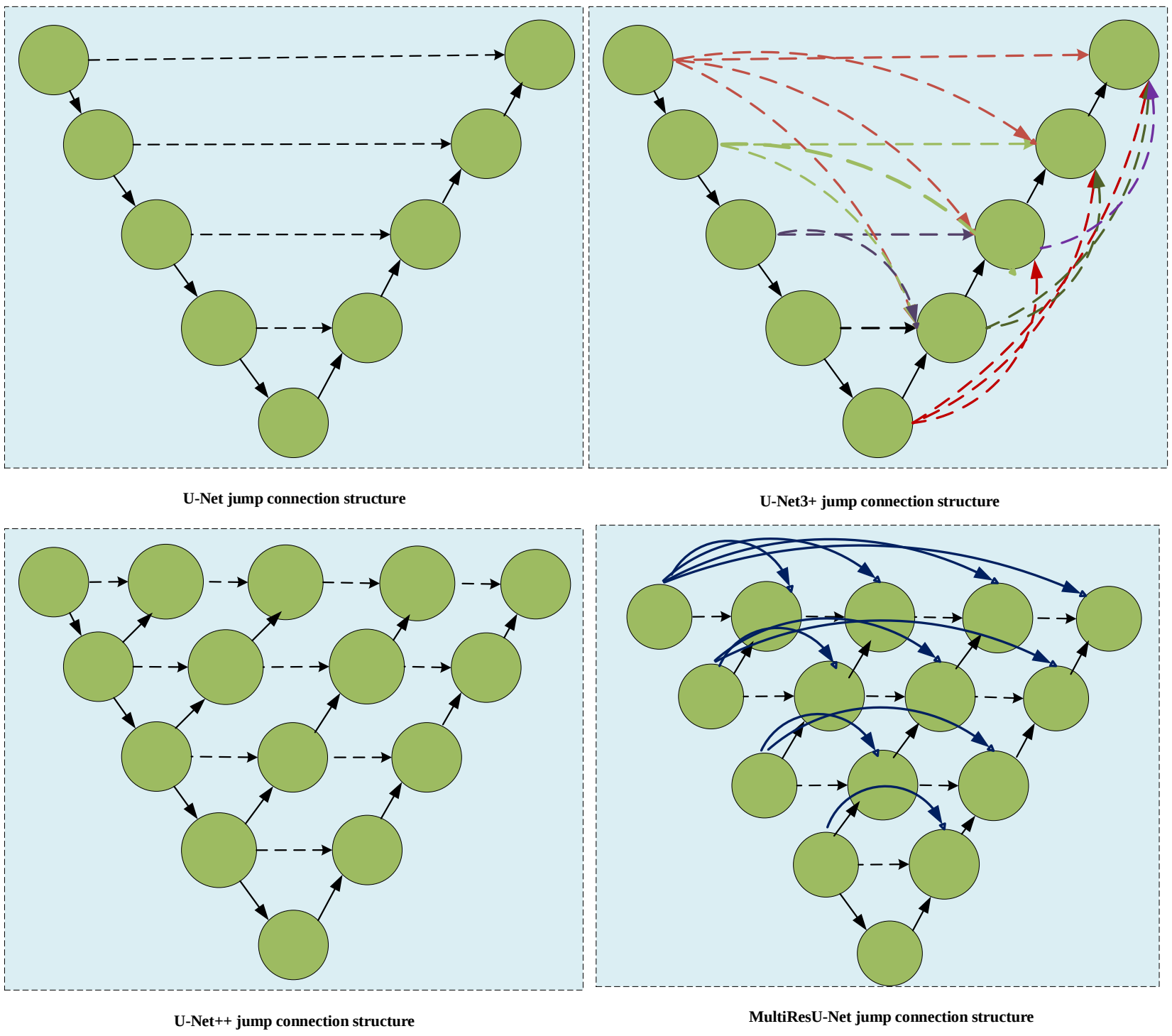

Figure 18.Based on U-Net four jump connection structure

### 3.2.3 Transformer mechanism.

To refine medical image segmentation accuracy for better clinical outcomes, semantic segmentation is effectively achieved by enriching contextual information, which enhances the distinction between patient lesions and surrounding tissues in medical images and elucidates the connections within individual image patches. The transformer architecture offers an optimal solution by utilizing a stacked encoder-decoder structure, effectively avoiding loops. The encoder is structured with two sublayers per layer: a multihead attention component and a feedforward connection component. Conversely, the decoder has three sublayers per layer: a masked multihead attention component, a multihead attention component, and a feedforward connection component. Every layer component is integrated with residual connections and regularization layers. The complete architecture is illustrated in Figure 19.

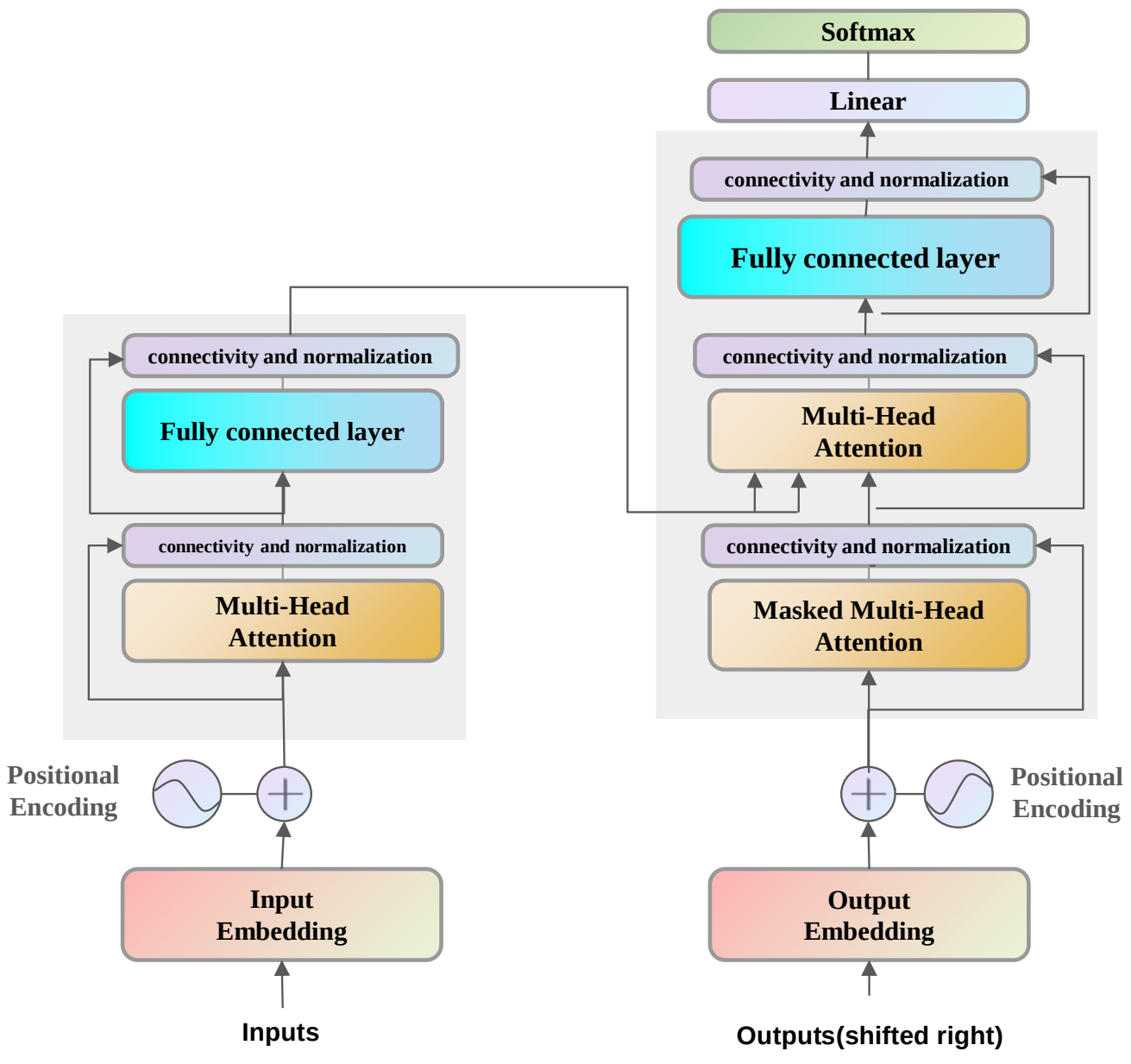

Figure 19. Transformer Model architecture

Multiple algorithms combine Transformer and U-Net to construct a network for lesion segmentation. Chen et al. introduced a TransUNet [38] that combines the strengths of U-shaped and transformers. Transformers encode and extract global top and bottom regions from annotated image blocks of feature maps from convolutional neural networks. With respect to the input sequence information of the text, the decoder performs samples on encoded features and is then combined with high-resolution CNN feature maps to achieve precise localization, assisting U-Net in recovering local spatial information and achieving precise segmentation in Synapse. The results of organ segmentation showed that TransUNet has better performance than U-Net does. The TransU-Net architecture is shown in Figure 20.

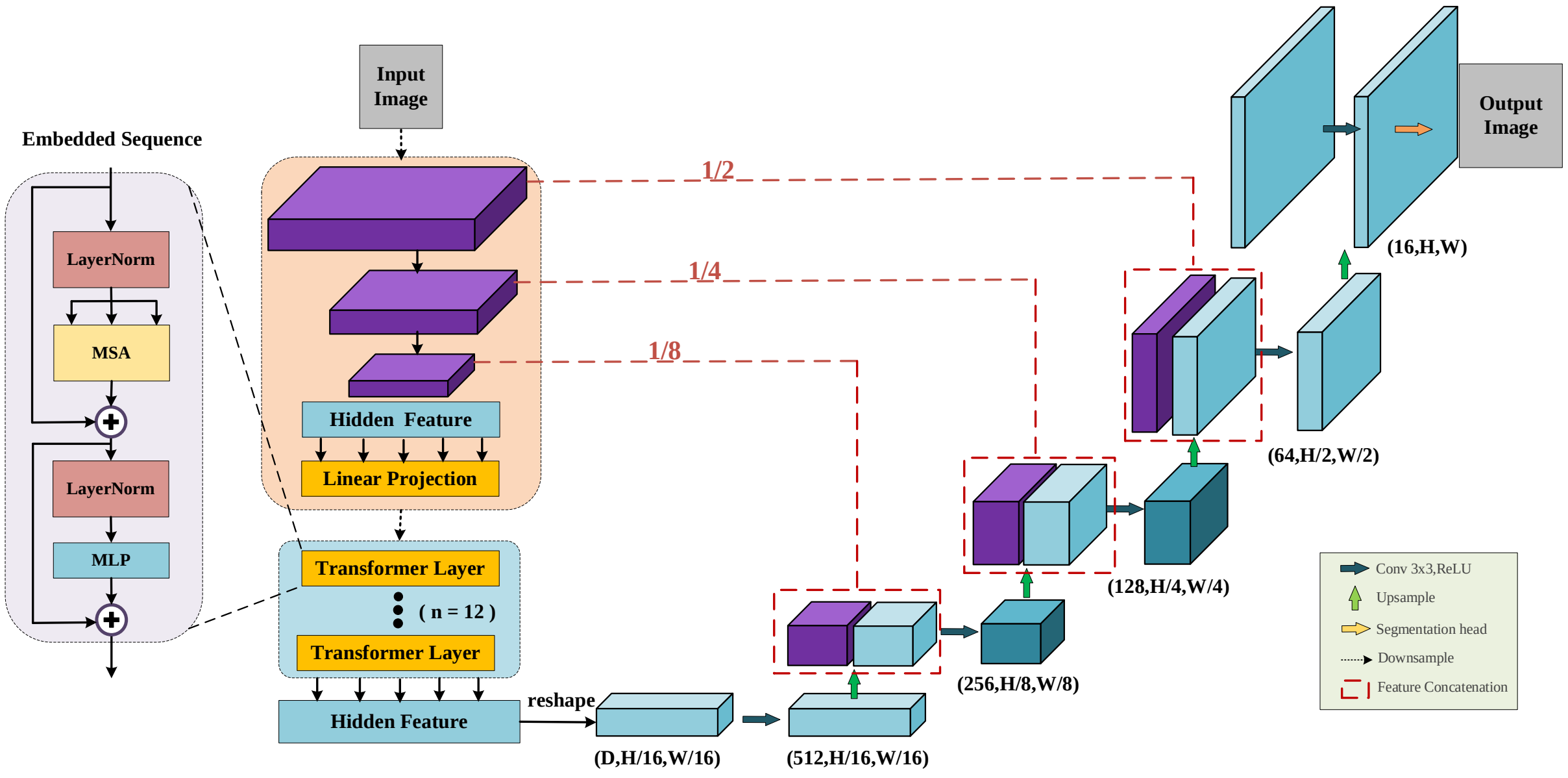

Figure 20. The Trans-UNet architecture (Chen et al.,2021)

In the tasks of colon polyp, skin, and prostate segmentation, Zhang et al. developed the TransFuse [39] model, which has a parallel branching architecture with two encoder branches: DeiT-S as the backbone branch to extract global information and ResNet as the backbone for the CNN branch to capture local information. The two encoder branches are used to perform the branching in the BiFusion module. TransFuse represents the first parallel branching model that merges a CNN with a transformer to overcome the problems of gradient vanishing and diminishing features. Figure 21 shows the details of the architecture of the Transfuse network.

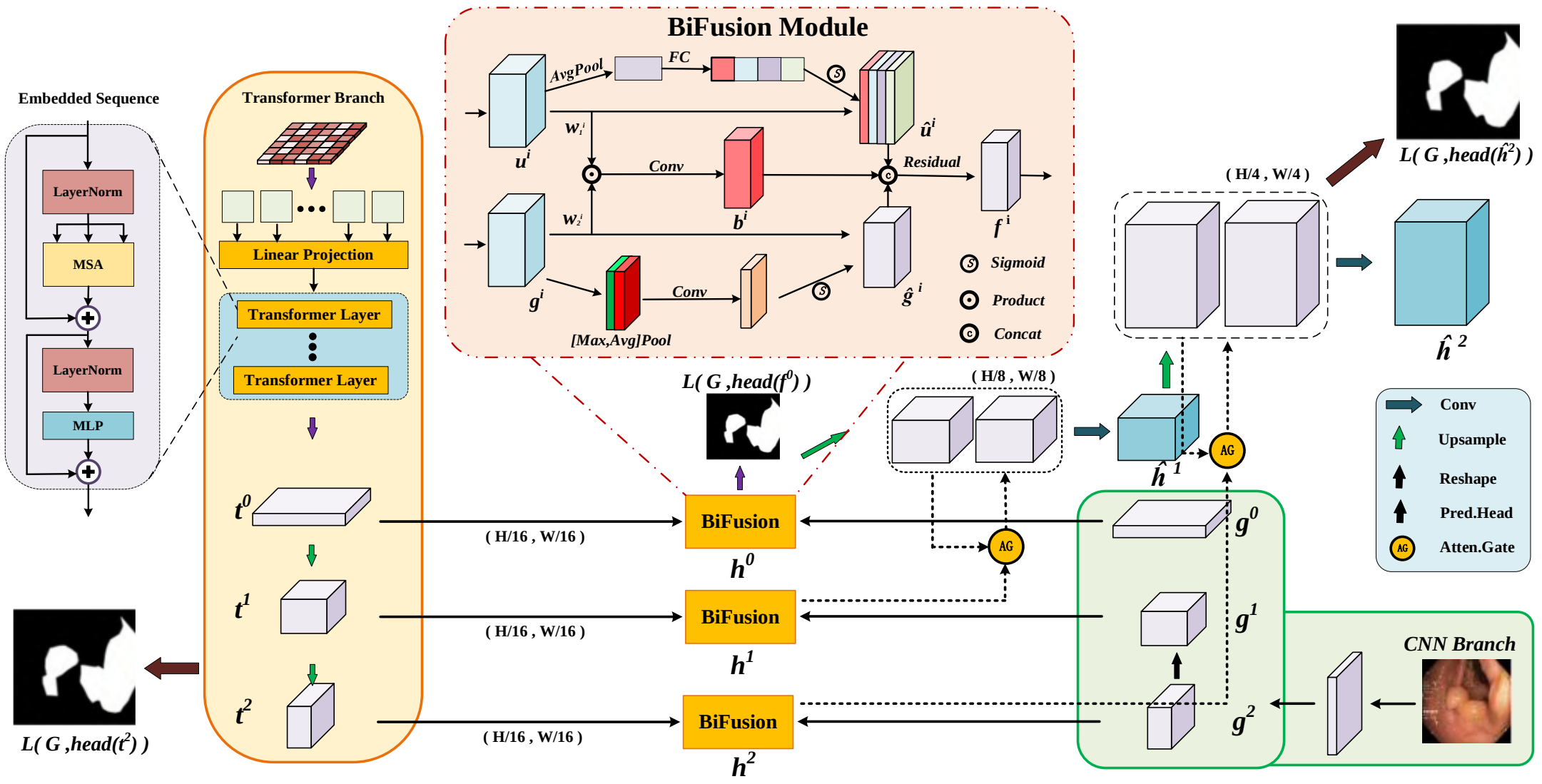

Figure 21. The Transfuse architecture (Zhang et al.,2021)

Cao et al. designed the transformer-based U-shaped network Swin UNet [40], which introduces the double transformer block as the encoder, decoder, and bottleneck, as shown in Figure 22, and the method demonstrates good segmentation accuracy and robust generalizability on the Synapse and ACDC segmentation datasets.

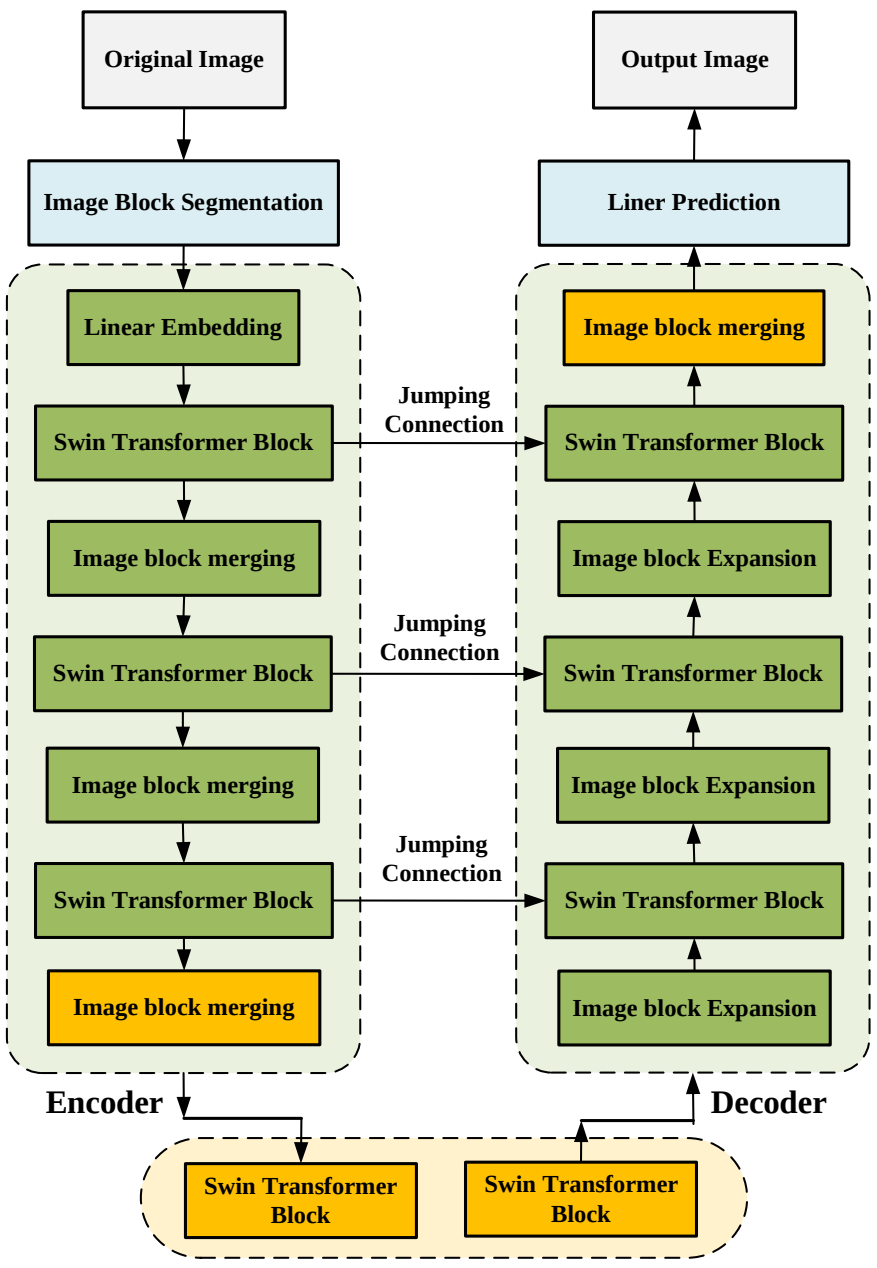

Figure 22. Swin-UNet architecture (Cao et al.,2023)

Libin Lan et al. propose BRAU-Net++[41], a U-shaped hybrid convolutional neural network (CNN-Transformer) architecture that incorporates a dual-path attention mechanism as its core building block. The network leverages dynamic sparse attention in place of conventional full attention or manually designed static sparse attention, enabling efficient learning of both local and global semantic information while significantly reducing computational complexity. Furthermore, a novel module, termed Skip-Connection Channel-Space Attention (SCCSA), is introduced to integrate multi-scale features, compensating for the loss of spatial information and enhancing cross-dimensional interactions.

The proposed method demonstrates state-of-the-art (SOTA) performance across nearly all evaluation metrics on the Synapse Multi-Organ Segmentation, ISIC-2018 Challenge, and CVC-ClinicDB datasets. Notably, it excels in capturing features of small targets, addressing a persistent challenge in medical image segmentation. In future research, the goal is to design more sophisticated and generalized architectures for multimodal medical image segmentation tasks to further improve the applicability and robustness of the proposed methods. Figure 23 shows the details of the architecture of the BRAU-Net++ network.

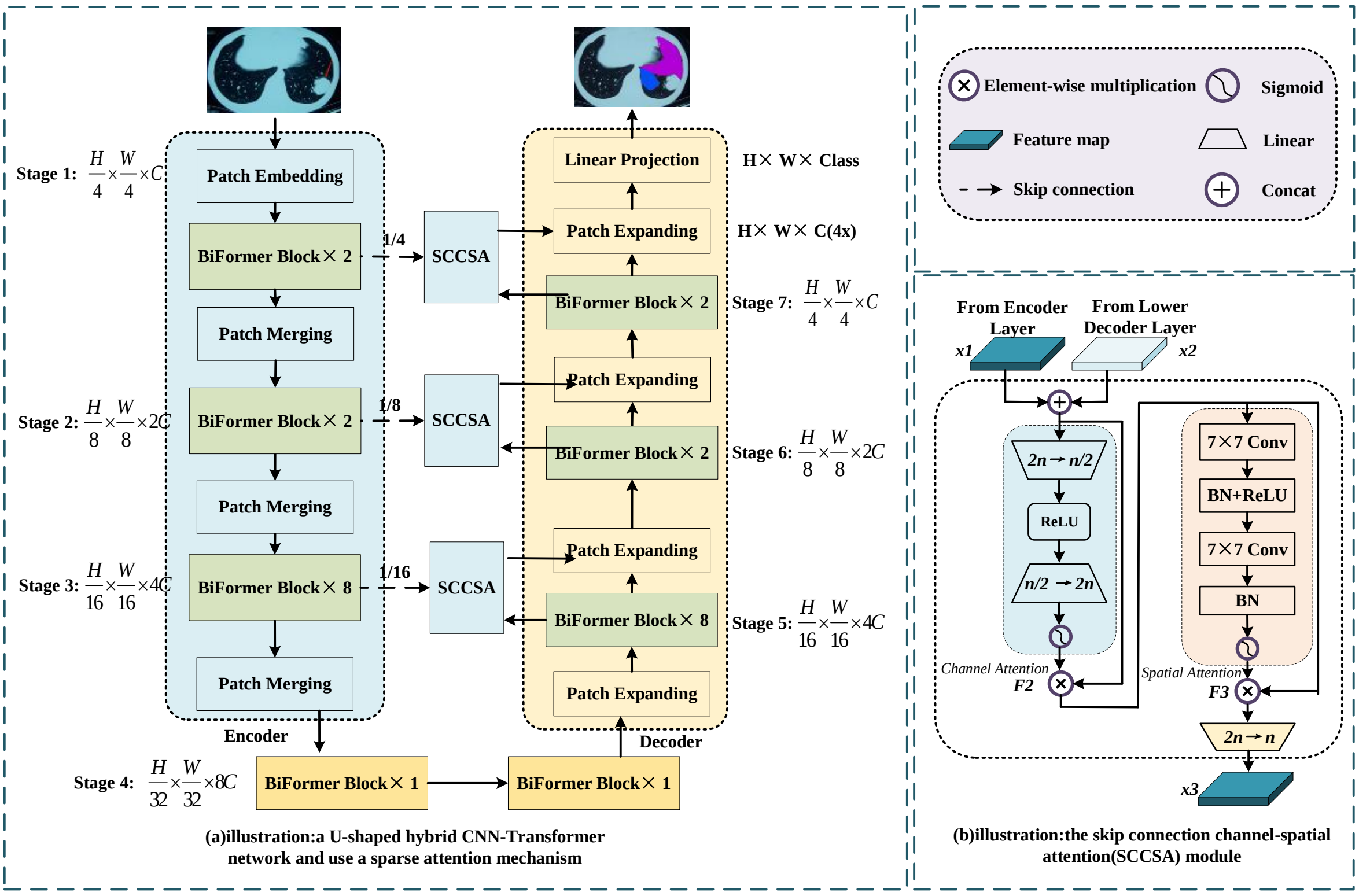

Figure 23 The Architecture of BRAU-Net++( Libin Lan et al.,2024)

### 3.2.4 Attention mechanism.

DCSAU-Net[42], proposed by Qing Xu et al., represents a novel encoder-decoder architecture designed for medical image segmentation. This model effectively addresses key limitations of traditional U-Net architectures, including insufficient feature extraction at varying depths and inefficiencies in handling complex medical images. The proposed framework is rigorously evaluated through comprehensive experiments and an in-depth ablation study, which elucidates the specific contributions of each component to the overall performance.

Despite its robust design and demonstrated efficacy, certain limitations remain. Extending the model to 3D segmentation and improving its capability to process images with high foreground-background similarity represent promising directions for future research. Overall, this work constitutes a significant advancement in the field of medical image segmentation, combining robust performance with computational efficiency. Further optimization and adaptation of the architecture could enhance its generalizability and applicability to a wider range of segmentation tasks. The complete architectural details are provided in Figure 24.

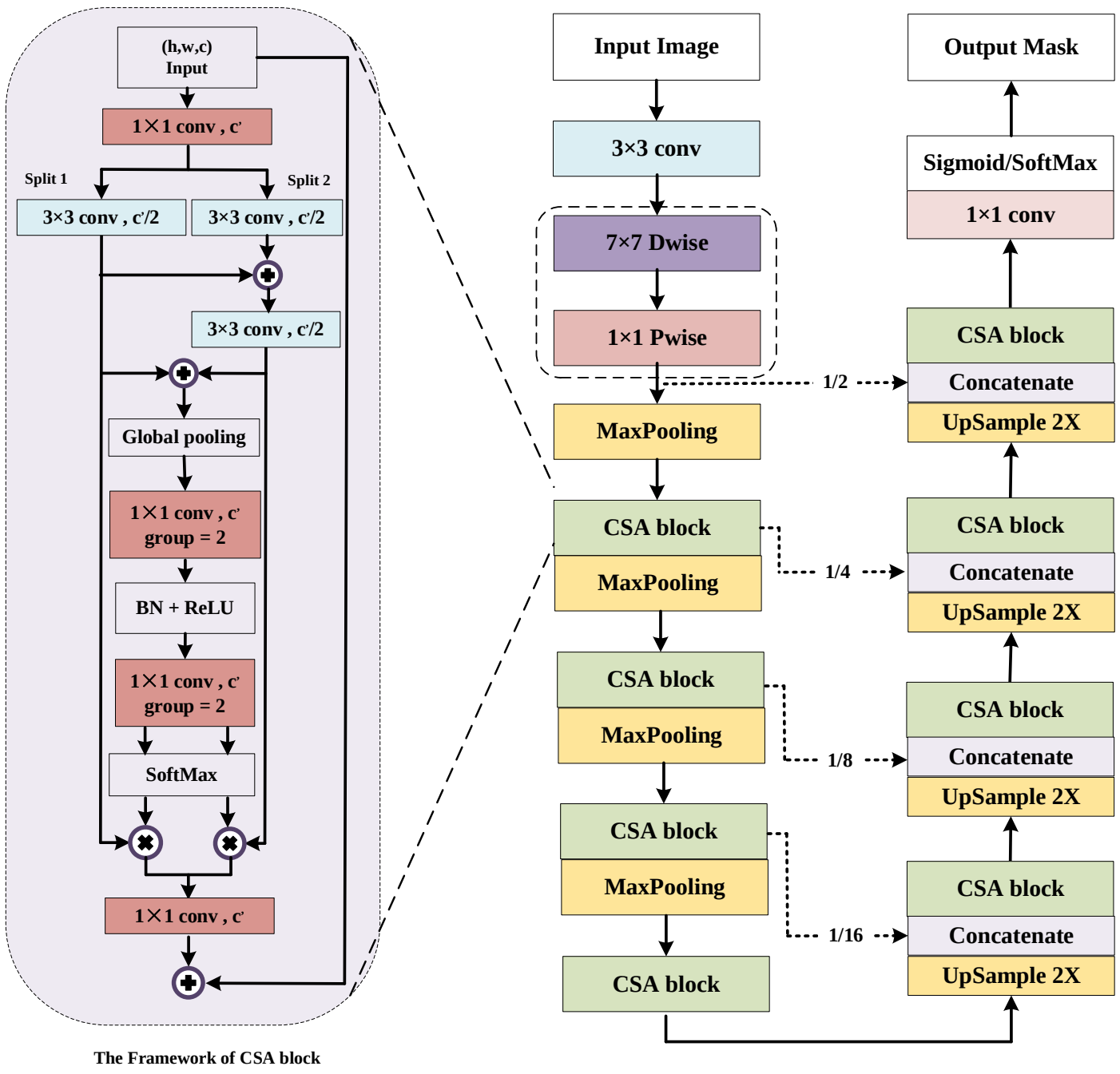

Figure 24 The architecture of DCSAU-Net(Qing Xu et al.,2023)

Md Mostafijur Rahman et.al introduces an efficient and innovative decoder architecture designed to address the computational challenges prevalent in medical image segmentation. By leveraging multi-scale depth-wise convolution blocks and integrating advanced attention mechanisms, including channel attention, spatial attention, and large-kernel grouped attention, the proposed EMCAD [43]framework enhances both accuracy and efficiency.

Empirical evaluations on 12 benchmark datasets across six segmentation tasks demonstrate EMCAD's superior performance. It achieves state-of-the-art results with notable reductions in computational complexity, requiring 79.4% fewer parameters and 80.3% fewer FLOPs compared to competing methods. These outcomes are attributed to the novel Multi-scale Convolutional Attention Module (MSCAM) and Large-kernel Grouped Attention Gate (LGAG), whose individual contributions are validated through comprehensive ablation studies. Despite its strengths, the current implementation focuses exclusively on 2D medical image segmentation, which may limit its applicability to 3D imaging tasks. EMCAD represents a significant advancement in medical image segmentation, balancing computational efficiency with high segmentation accuracy. Its modular and adaptable design positions it as a promising tool for broader applications in medical image analysis. The complete architectural details are provided in Figure 25.

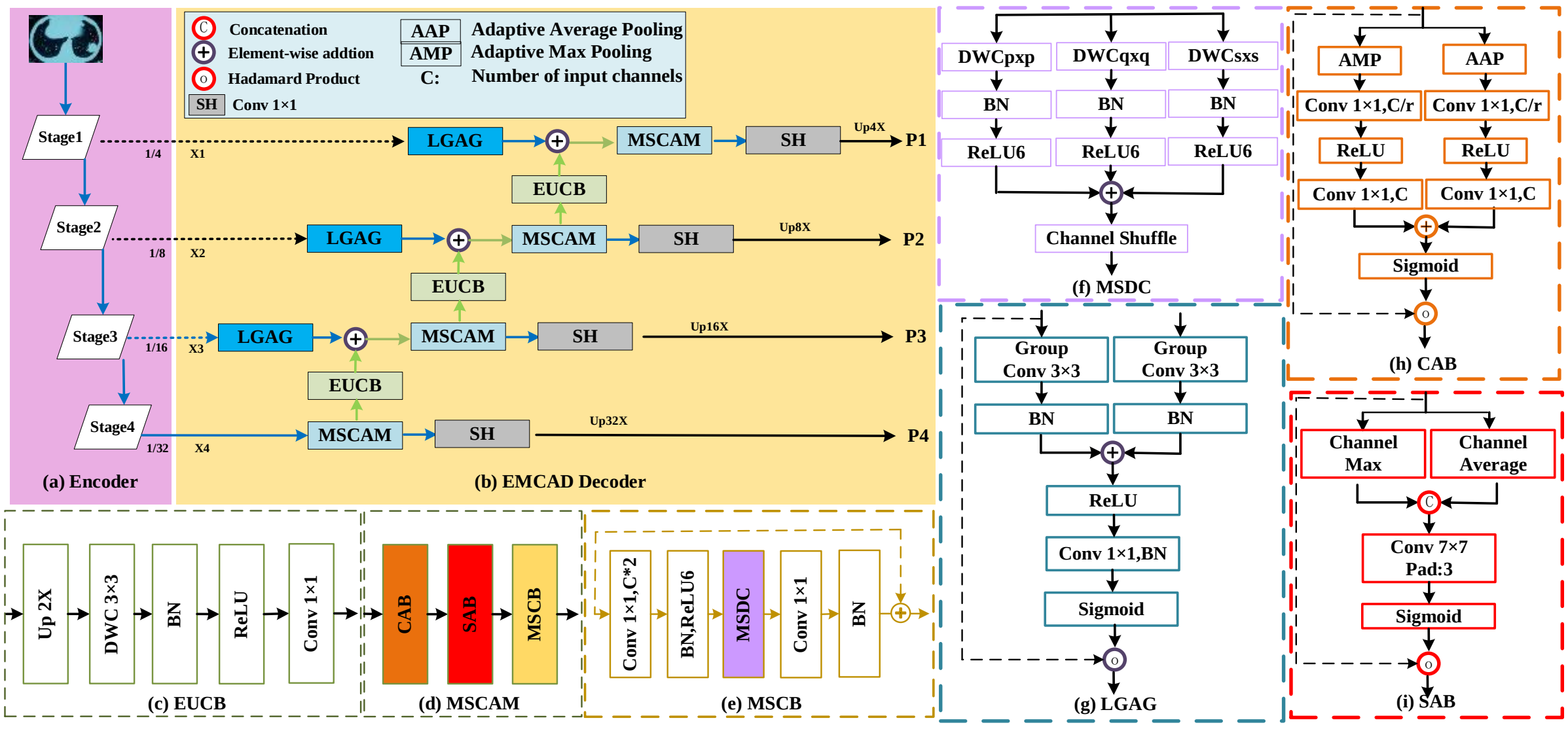

Figure 25.The architecture of EMCAD Decoder (Md Mostafijur Rahman et.al.,2024)

## 4 Evaluation Metrics.

The selection of appropriate evaluation indices depends on specific application scenarios and needs; for example, if the concern is boundary accuracy, the Hausdorff distance or average surface distance may be more appropriate choices; if the concern is overall segmentation accuracy and consistency, the Dice coefficient or Jaccard index may be more applicable. In practical applications, it is often necessary to consider various metrics to evaluate a model's performance fully.

### 4.1 Dice Similarity Coefficient (DSC)

The DSC is a statistical metric that measures similarity by the ratio of intersection and union. Its value range is between 0 and 1, where 1 indicates that the two images are completely identical, expressed by the following formula:

$$DSC = \frac{2\left|Area_{pre} \cap Area_{tru}\right|}{\left|Area_{pre}\right| + \left|Area_{tru}\right|}$$

### 4.2 Jaccard Index (JI)

Similar to DSC, but with a slightly different calculation method, it focuses on measuring the predicted segmentation area ( $Area_{pre}$ ) that matches the ground truth segmentation area ( $Area_{tru}$ ).

$$JI = \frac{\left|Area_{pre} \cap Area_{tru}\right|}{\left|Area_{pre} \cup Area_{tru}\right|}$$

### 4.3 Mean Intersection over Union (MIoU)

MIoU is a widely used evaluation metric for assessing the performance of a model in segmentation. The IoU, first, calculates the proportion of overlapping and combined parts via the following formula:

$$IoU = \frac{Area_r \cap Area_p}{Area_r \cup Area_p}$$

where $Area_r$ represents the real image area and where $Area_p$ represents the predicted image area.The MIoU represents the average IoU across all categories. Assuming that there are m categories, the formula for the MIoU is as follows:

$$MIoU = \frac{1}{m}\sum_{i=1}^{m} IoU_i$$

### 4.4 Hausdorff Distance (HD)

In measuring the effectiveness of the segmentation of medical datasets, HD is usually used to indicate the accuracy of the predicted segmentation relative to the ground truth segmentation. The value of HD reflects the degree of coincidence between the predicted contour and the real contour of the segmented object. The formula for HD is as follows:

$$d_H = \max\left\{\sup_{a\in A}\inf_{b\in B} d(a,b), \sup_{b\in B}\inf_{a\in A} d(a,b)\right\}$$

### 4.5 Confusion matrix-related indicators

| | **Actually Postive** | **Actually Negative** |
|---|---|---|
| **Predicted Postive** | ***TP*** | ***FP*** |
| **Predicted Negative** | ***FN*** | ***TN*** |

**T_P:**The model predicts correctly(True) and predicts as a positive sample(Positive)
**F_P:**The model predicts error(False) and predicts as a positive sample(Positive)
**F_N:**The model predicts error(False) and predicts as a Negative sample(Negative)
**T_N:**The model predicts correctly(True) and predicts as a Negative sample(Negative)

Figure 26 Confusion matrix

- **Accuracy:**
  Accuracy indicates the number of samples that can be correctly predicted as a proportion of the total number of samples.

$$Accuracy = \frac{TP + TN}{TP + TN + FP + FN}$$

- **Recall (Sensitivity)**

Recall that, also known as sensitivity, it narrows the range of sample ratios to positive samples and calculates the ratio of correctly predicted samples to actual samples. In medical diagnostic applications, where sensitivity to the risk of false negatives is critical, recall is an essential evaluation metric. Recall is defined as:

$$recall = \frac{TP}{TP + FN}$$

## 5. Challenges and Solutions

### 5.1 Limitations of medical datasets

Most segmentations for medical datasets rely on supervised deep neural network models, and high-quality medical image segmentation annotation requires expert-level knowledge, which demands a significant amount of time and incurs high costs. Additionally, manual annotation has difficulty avoiding subjective errors. Models based on the ViT architecture require sufficient samples to establish global relationships, and most medical image datasets are not of a size that can satisfy the pretraining requirements. This is usually solved by data enhancement methods, such as elastic deformation [10] designed by Ronneberger et al., which generates a smooth deformation of the original image due to the random displacement vector; Beji et al. proposed Seg2GAN [44], for synthesizing data distributions for data enhancement; Milletari et al. proposed V-Net [3] to input arbitrarily transformed nonlinear images into a dense deformation field; and; Wu et al. proposed the D-former [45] expansion concept, where the self-attention mechanism captures feature information, enabling the extraction of global information even with a small sample size.

### 5.2 Insufficient model generalization.

In medical image segmentation, the ratio of the region of interest to the whole image is usually quite small, which is reflected in the extreme imbalance between positive and negative samples. In general, it is difficult to maintain stable performance of the same network model for the segmentation of different medical image datasets. Migration learning can be performed by fine-tuning partial parameters of the pretrained mode to adapt it to other datasets, but labeled data cannot be avoided. Consequently, collecting substantial amounts of data from new domains to retrain the network incurs significant costs. Yan et al. proposed that Unet-GAN [46] externally accesses the Cycle-GAN network to change the feature distribution in the current data to adapt to the distribution of the original domains to ensure segmentation accuracy across datasets; Huang et al. proposed that 3D U2-Net [47] applies domain adapters inside the codecs for depth-separable convolution to adapt to different datasets to achieve multiorgan segmentation; Isensee et al. proposed nnU-Net [7], which collects the target dataset features to design the segmentation algorithm, adjusts various hyperparameters, and improves model generalization performance. Unsupervised learning methods applied to the field of medical image dataset segmentation can be divided into two categories: domain adaptation and domain generalization. These methods achieve consistency in the feature space and category space of the source and target domains while ensuring a consistent distribution of features. Because these two methods still require large-scale data labeling, the development of unsupervised learning models with small amounts of data labeling has become an important future direction for medical image segmentation models.

### 5.3 Weak Robustness and Poor Stability.

In medical image segmentation, segmentation models are required to maintain good performance and effectively improve robustness and stability in the face of blurred, low-contrast, and shadowed medical images. A stable and robust model can provide more accurate segmentation results and support doctors in developing more accurate diagnostic and treatment plans, thus improving the quality and safety of patient care. Improving the stability and robustness of the U-Net model starts from these three aspects.

### 5.3.1 Regularization

- Regularization: A weight penalty term is added to the loss function to avoid model overfitting.
- Dropout: This weakens the interdependence between neurons by randomly discarding a portion of the neurons to improve the generalization ability of the model.

### 5.3.2 Loss function optimization

- Dice loss: For the segmentation task, dice loss can better handle the unbalanced foreground and background situation and improve the segmentation accuracy.
- Focal Loss: Based on cross-entropy, a higher weight is given to difficult classification samples to help the model focus on learning difficult samples.

### 5.3.3 Training Strategies

- Learning Rate Scheduling: Dynamic adjustment of learning rates, such as cosine annealing and exponential decay, to promote the model's stable convergence.
- Early Stopping Method: The real-time monitoring model terminates training at the point where the performance curve tends to flatten during validation of the dataset, avoiding overfitting.

The proposed solution in this study combined with the study of the problem of scarce data for medical image annotation does not completely eliminate this fundamental drawback and may limit the development of deep network models. However, many scholars' improvements can enhance model performance with limited data. The advantages and disadvantages of the improvement work of each network structure are compared, as shown in Tables 4~6.

Table 4. Based on the skip connection module application

| Model | Time | Dataset | 2D/3D | Evaluation Indicator | Values | Advantage | Disadvantage | Improvement Methods |
|---|---|---|---|---|---|---|---|---|
| Attention U-Net++[48] | 2020 | MICCAI-2017 | 2D | DICE | 98.15% | High accuracy and Specificity | High computational costs. | Lightweight Attention and Hybrid Training |
| RA-Unet[49] | 2020 | LiTS/3DIRCADb | 3D | DICE | 96.10% | Strong feature capture and high accuracy | High computational costs. | Model pruning and Expand Applicability |
| Attention 3D U-Net[50] | 2022 | BRATS-2020 | 3D | DICE | 89.74% | | | |
| MultiResU-Net[32] | 2020 | Codella et al.2019 | 2D | Jaccard Index | 91.65% | High accuracy and Specificity | | |
| DCSAU-Net[42] | 2023 | ISIC2018 | 2D | DSC | 90.40% | | Limited Flexibility and overfitting | |
| TransUNet+[37] | 2022 | Synapse/ACDC/GlaS | 2D | DSC | 90.47% | Transformers Integration | High computational costs. | Model Simplification and data augmentation |
| FSC-UNet[51] | 2022 | Cell-nuclei<br>Heart<br>Chaos | 3D | IOU | 96.7%<br>95.6%<br>95.9% | High accuracy and efficiency | Limited Flexibility and overfitting | |
| U-Net3+[35] | 2020 | LiTS-2017 | 3D | DICE | 96.75% | | Computationally intensive and overfitting risk | |

## Table 5. segmentation application based on Transformer module

| Model | Time | Dataset | 2D/3D | Evaluation Metrics | Values | Advantage | Disadvantage | Improvement Methods |
|---|---|---|---|---|---|---|---|---|
| TransUNet[38] | 2021 | Synapse<br>ACDC | 3D | DICE | 77.48%<br>89.71% | Enables global modeling | High computational complexity | Based on CNN |
| TransBTS[52] | 2021 | BraTS-2019 | 4D | DICE | 90.01% | Full 3D image input possible | - | |
| CoTr[53] | 2021 | Synapse | 3D | DICE | 85.00% | 3D Residual Convolution Block | - | |
| UNETR[25] | 2021 | BCTV<br>MSD | 3D | DICE | 89.1%<br>96.4% | Learns global multi-scale features of the 3D image | Insufficient model adaptation capability | |
| TransU2-Net[54] | 2023 | BraTS-2021 | 5D | DICE | 88.17% | Context aware mechanism automatically adjusts global relationships | - | |
| UCTransNet[55] | 2022 | MoNuSeg | 2D | DICE | 90.18% | Parallel encoder branching fuses multilevel features | Loss of information during upsampling | |
| MedT[56] | 2021 | MoNuSeg | 2D | F1/Iou | 88.84%<br>81.02% | Visualization of the attention map | Limited model performance | Based on Transformer |
| UNETR[25] | 2022 | BCTV<br>MSD | 3D | DICE | 89.1%<br>96.4% | Channel Transformer eliminates semantic gaps between codec features | - | |
| MISSFormer[57] | 2023 | Synapse<br>ACDC | 3D | DSC | 76.10% | Adaptive window generation at object boundaries to mitigate boundary ambiguity | - | |
| Swin-UNet[40] | 2023 | ACDC | 2D | DSC | 79.13% | Gated axial attention mechanism controls information flow | Lack of multi-scale information representation | |
| D-Former[45] | 2022 | Synapse<br>ACDC | 3D | DSC | 88.83%<br>92.29% | Efficient self-attention module reduces attention complexity | Lack of multi-scale information representation | |
| Dilated-Unet[58] | 2023 | Synapse<br>ACDC | 3D | DSC | 89.92%<br>94.26% | Realizes local-to-global self-attention | Inability to handle spatial information | |
| Swin-UNETR[59] | 2022 | BraTS-2021 | 3D | DICE | 92.70% | Swin Transformer block builds dual-scale encoder | High computational complexity | Based on Hybrid Coding |
| nnFormer[60] | 2023 | Synapse<br>ACDC | 3D | DICE | 90.10% | Expansion method for self-attention to local and global feature relations to control computation efficiently. | - | |
| CS-UNet[61] | 2023 | BraTS2019-2021 | 3D | DICE | 83.32% | Design Dilated block to realize sparse global attention. | Large number of parameters, high computational complexity | |
| UNETR++[62] | 2023 | Synapse<br>SegPC-2021 | 3D | DSC | 80.6%<br>92.3% | Combining multi-scale contextual representation and 3D image processing | - | |
| BRAU-Net++[41] | 2024 | ISIC-2018<br>CVC-CLINICDB | 3D | DSC | 90.1%<br>92.94% | Good global feature capture capability and high segmentation accuracy | Limitations on generalization ability | |

## Table 6. Segmentation application based on the 3D U-Net Network

| Model | Time | Dataset | 2D/3D | Evaluation Metrics | Values | Advantages | Disadvantages | Improvement methods |
|---|---|---|---|---|---|---|---|---|
| PGL[63] | 2021 | RibFrac/KiTS | 2D | DICE | 95.60%<br>84.29% | Good global feature capture capability and Strong generalization | Insufficient segmentation performance | Use weight-sharing mechanisms to enhance multi-task learning. |
| UNETR[25] | 2022 | Synapse<br>BTCV<br>ACDC | 2D/3D | DSC | 87.2%<br>83.28%<br>92.83% | Good global feature capture capability and high segmentation accuracy. | High computational costs. | Optimize network architecture and data augmentation |
| MA-Net[64] | 2021 | MICCAI-2017<br>LiTS | 4D | DICE | 96.00% | Strong feature extraction capability | Limitations on generalization ability | |
| LKAU-Net[65] | 2021 | CT-ORG<br>BraTS-2020 | 3D | DICE | 92.15%<br>84.81% | Strong generalization and feature extraction. | computational efficiency details are unclear. | |
| LeViT-UNet[66] | 2021 | Synapse<br>ACDC | 2D/3D | DSC | 78.53%<br>90.32% | High segmentation accuracy and efficiency | High computational costs. | |
| DiNts[67] | 2022 | MSD | 3D | DICE | 80.20% | Strong generalization and high accuracy | | |
| 3D MCSE-Net[68] | 2022 | TPCTS<br>MICCAI-2017 LiTS | 3D | DICE | 76.11%<br>73.7% | High specificity and sensitivity | | |
| CLD-Net[69] | 2022 | Kvasir-SEG<br>CVC-ClinicDB | 2D | DSC | 93.1%<br>94.64% | High accuracy | Limitations on generalization ability | |
| DSCA-Net[70] | 2023 | ISIC 2018<br>TNBC<br>LUNA | 5D | DICE | 92.82%<br>97.27%<br>98.28% | Strong generalization and high accuracy | High computational costs. | Implement model pruning or quantization |
| COSMOS[71] | 2022 | MICCAI-2021<br>crossMoDA | 3D | DICE | 87.1%<br>84.2% | High spatial resolution | Limitations on generalization ability | Data augmentation |
| UNETR++[62] | 2023 | MSD | 3D | DICE | 87.22% | High accuracy and robustness | | |

As demonstrated in Tables 4 to 6, various U-Net variants present the following challenges in medical image segmentation:

(1) Limited capacity to capture global features: In U-Net and its variants, the convolutional layer serves as the network's backbone, inherently restricting the receptive field owing to its local connection mechanism. This mechanism limits the ability of the model to capture global features, consequently affecting segmentation accuracy. This issue is evident in models such as H-DenseUNet[5] and UNet++[72].

(2) High computational consumption. Medical images are high in resolution and rich in semantic information; CNN models need to build deep network structures for multiple convolutions of each pixel, increasing the parameter magnitude and network structure complexity of the model, such as V-Net[3], CE-Net[73], Dense-Inception U-Net[74], and W-Net[18]. On the basis of the model architecture, improvement will be incorporated into the performance module or the combination of additional network branches; although it can reduce the workload of labeling datasets, it also leads to model complexity, so the computational complexity increases. For example, TransUNet[38]TransBTS[52],TransFuse[39],nnFormer[60] and RA-UNet[49].

(3) Insufficient generalization capacity. Differences in neighborhood distributions among different datasets result in models trained on a specific dataset not generalizing well to other neighborhoods, affecting the model’s generalization performance. For example, Cascaded Bionet[36] and FECC-Net.[75]

(4) Insufficient boundary localization ability: Because medical images contain much spatial information and the ViT architecture model focuses mainly on the data relationships between sequences, it is not sensitive to fine-grained features and has difficulty capturing the spatial relationships between pixels, leading to weak boundary localization ability. Examples include TransUNet [38], CoTr[53], TransFuse [39], CS-net[76], and Swin-Unet[40].

To effectively improve the deficiencies in the above models, this paper proposes the following reference suggestions for different research purposes:

(1) Enhanced feature extraction. Adding residual or dense connections between convolutional blocks improves deep feature extraction, e.g., V-Net[3], Res-UNet[77], MultiRes[32], and FD-Unet[78]; jump connections combined with the dense idea can maximize the fusion of semantic features at different scales, e.g., UNet++[72], Net3+[35], and UNet-Sharp; and hybrid network subblocks provide sufficient receptive fields for 3D image learning and facilitate the extraction of local to global features, e.g., hybrid network subblocks provide enough sensory fields for 3D image learning and facilitate the extraction of local to global features, e.g., informers, UNETR++ [32], and PHTrans[79].

(2) Expanding the receptive field. Expansive convolutional substitution network submodules, such as dense multipath U-Net[17]; adding encoder branches; and integrating ASPP, such as M-Net[23], connected UNets, and layer-by-layer expansion of the perceptual field by using layer-by-layer connectivity and multiscale feature fusion, for example, Res-UNet[77], help increase the perceptual field. The sensory field can specialize in capturing long-distance features via the transformer self-attention mechanism.

(3) Enhanced spatial information extraction: TransFuse can capture features and fuse them effectively through parallel two-branch design, which provides high flexibility and adaptability when processing different types of medical images. BATFormer[80] is specially designed with a boundary-aware module, which establishes multiscale feature mapping to obtain global feature localization information and has high segmentation accuracy. UNETR++[81] introduces a multiscale feature fusion strategy into the transformer network structure to optimize the deep network model, thus enhancing information extraction in both spatial and channel dimensions.

(4) Reduce the computational complexity. Using depthwise separable connections, densely connected networks implement feature reuse to increase information flow while occupying fewer parameters, e.g., FD-Unet[78] and MDU-Net[82]; inhibit network complexity by pruning the network structure and parameter quantization, e.g., Dense-Inception U-Ne[74]t and Inception-Res; effectively reduce computational effort by sharing weights to encode the spatial and channel-dimensional information to reduce computation effectively, e.g., UNETR++ [81]; and low-cost mining of global dependency information by self-attention to captured local and global feature relationships to reduce computation, e.g., D-Former[45].

## 6. Future Directions.

In recent years, U-Net and variant network models have shown excellent performance in medical dataset segmentation. With advancements in GPU computing and their increasing applications, segmentation models and transformers have been combined, and the model's segmentation accuracy and overall performance have greatly improved. This study proposes the following points for future research directions in light of the current state of research. This paper provides a comprehensive analysis of U-shaped network models for segmentation, focusing on an improvement mechanism based on jump connections, residual connections, 3D U-Nets, and transformers. The strengths, weaknesses, and challenges of its various enhancement methods are shown in Table 7.

Table 7 Strengths and weaknesses of the four improvement methods and challenges

| Improvement methods | Strengths | weaknesses | Improvement direction | Facing challenges |
|---|---|---|---|---|
| **Jump connection mechanism** | ●Prevent information loss<br>●Reduce gradient dispersion<br>●Accelerate model convergence | ●Information redundancy<br>●Same scale feature map fusion | ●Increase the number of connections<br>●Jump connections introduce attention<br>●Jump connections introduce feature fusion | ●Improve network segmentation accuracy<br>●Overcome noise interference |
| **Transformer mechanism** | ●Capture contextual dependencies<br>●Pre training and adjust slightly | ●Modeling local visual features lacks<br>●Lack of inherent inductive bias<br>●Poor interpretability | ●Transformer combined with U-Net<br>●Improvements based on Swin Transformer | ●Solve the problem of low contrast at lesion edges<br>●Overcome the phenomenon of unclear presence of individual image block levels |
| **3D U-Net** | ●Mining data for high-dimensional correlation<br>●Extracting deep image features | ●High computational cost<br>●GPU memory consumption | ●Attention Mechanism Improvement<br>●Self-supervised 3D U-Net<br>●Encoder and decoder improvements | ●Data annotation of labor cost inputs |
| **Residual Connection Mechanism** | ●Resolving vanishing gradients<br>●Overcoming model degradation<br>●Powerful feature characterization | ●High model complexity<br>●Large number of parameters<br>●Feature distortion problem<br>●Poor generalization ability for small-scale data | ●Improved Convolutional Layer<br>●Recursive residual units<br>●Attention residual unit | ●Enhanced model robustness<br>●Increased utilization of feature information<br>●Solve the fuzzy noise boundary problem |

In the medical dataset segmentation task, challenges such as class imbalance and low contrast are predominantly addressed by augmenting the foundational U-Net architecture with multiscale modules and multifeature fusion mechanisms. Additionally, to mitigate information loss, many approaches incorporate skip connections and shared network parameters, thereby increasing the accuracy of pixelwise classification. To improve the underutilization of information in multimodal medical image segmentation, making the target area more prominent and conspicuous via medical image data fusion methods can provide the network with more comprehensive and more detailed information about the pathology or organs so that the network can obtain better expression ability and judgment ability. The four improvement mechanisms summarized comprehensively address the limitations of the medical image segmentation task, and these studies have contributed to the development of medical image segmentation. Despite the remarkable results achieved, there is still much room for improvement in the field of medical image segmentation.

### 6.1 Interpretability of the model

The difficulty of deep neural network models in interpreting their output makes it highly important to explain segmentation from a clinical perspective. By visualizing maps of features and decision processes at different levels, we help physicians understand the model output labels, thus gaining insight into the label generation process and predicting the lesion trends through salient maps and explaining the reasons for the generation of lesions and the diffusion process. Physicians first gain a comprehensive understanding of the working mechanism of the model and then use their expertise to guide the design of the network architecture and optimize the decision-making process, thereby improving the segmentation performance of the model.

### 6.2 Segment Anything Model (SAM)

This model is a broad-based approach to image segmentation that effectively handles diverse segmentation tasks. Its application has excellent potential but needs further optimization and research. We can start with the following aspects:

- The SAM is optimized for medical image characteristics to adapt to high-resolution and complex structures.
- The segmentation accuracy on different datasets can be improved through fine-tuning and migration learning.
- Boundary perception and detail processing capabilities should be enhanced.
- The computational efficiency is optimized to achieve real-time segmentation.
- SAM can be integrated into the clinical system, intelligent assisted diagnostic tools can be developed, and the diagnostic accuracy and efficiency can be improved.

### 6.3 Feature Characterization Enhancement

From the perspective of feature representation, accurate modeling of the local structure and global context is crucial for medical image segmentation tasks, especially when 3D images are involved. With the implementation of a multiscale fusion mechanism in the network or the use of weakly supervised algorithms for feature extraction and ROI region segmentation, the data labeling problem can be solved, feature representations can be enriched, and the overall contextual information of the image can be better captured, which will help the model comprehensively comprehend the structures in medical images.

## 7 Conclusion

The greatest challenge in medical image analysis and processing has long been image segmentation. Accurate medical image segmentation is the cornerstone of 3D visualization of pathological tissues, surgical simulation and image-guided surgery. This is the most critical aspect of achieving an accurate diagnosis and developing an optimal treatment plan. This study comprehensively reviews the U-Net model and its four major variants, evaluating their performance across various medical imaging modalities and datasets. Furthermore, it delves into the application contexts, architectural compositions, and performance enhancements of these U-Net-based variants, facilitating a thorough understanding of the development directions and improvement strategies of U-Net models. Despite the ongoing challenges in image analysis, the U-Net architecture holds significant innovative potential and value in the analysis and processing of medical datasets. Advancements in network designs based on the U-Net framework are likely to result in significantly improved performance in the future.